\documentclass[12pt]{article}

\usepackage[a4paper,left=30mm,right=20mm,top=20mm,bottom=20mm]{geometry}
\usepackage{graphics}
\usepackage{amsmath}
\usepackage{amssymb}
\usepackage{epsfig}
\usepackage{cite}
\usepackage{xcolor}
\usepackage[unicode]{hyperref}
\title{Towards the global fit of the TMD gluon density in the proton from the LHC data}

\author{A.V.~Lipatov$^{1,2}$, G.I.~Lykasov$^{2}$, M.A.~Malyshev$^{1}$}

\begin{document}

\maketitle

\begin{center}

{\it $^{1}$Skobeltsyn Institute of Nuclear Physics, Lomonosov Moscow State University, 119991, Moscow, Russia}\\
{\it $^{2}$Joint Institute for Nuclear Research, 141980, Dubna, Moscow region, Russia}\\

\end{center}

\vspace{0.5cm}

\begin{center}

{\bf Abstract }

\end{center}

\indent

We propose a new analytical expression 
for the Transverse Momentum Dependent (TMD, or unintegrated) gluon density in the proton.
Essential phenomenological parameters
are extracted from the 
LHC data on inclusive hadron production in $pp$ collisions at low transverse momenta,
$p_T \sim 1$~GeV. The latter are described in the framework of modified soft quark-gluon string model, 
where gluonic state and non-zero transverse momentum of partons inside the proton
are taken into account.
To determine the parameters important at moderate and large $x$
we used
measurements of inclusive $b$-jet 
and Higgs boson production at the LHC
as well as latest HERA data on 
proton structure functions $F_2^c(x, Q^2)$ and $F_2^b(x, Q^2)$
and reduced cross sections $\sigma_{\rm red}^{c}(x,Q^2)$ and $\sigma_{\rm red}^{b}(x,Q^2)$.
The Catani-Ciafaloni-Fiorani-Marchesini evolution equation
is applied to extend the initial gluon distribution into the whole kinematical region.
We have achieved simultaneous description of all considered
processes with $\chi^2/d.o.f. = 2.2$,  
thus moving forward to the global fit of TMD gluon density from collider data.
The obtained TMD gluon distribution in a proton is available for public usage and 
implemented into the \textsc{tmdlib} package and 
Monte-Carlo event generator \textsc{pegasus}.


\vspace{1cm}


\newpage

\section{Introduction} \indent

It is known that
theoretical description of 
any physical observables, 
measured in the collider experiments, is mainly based on different factorization theorems
in Quantum Chromodynamics (QCD). These theorems provide the necessary framework to separate 
hard partonic physics, described with the perturbative QCD expansion, from soft hadronic physics,
described in terms of parton density functions (PDFs). The latter contain 
information on the non-perturbative structure of a hadron (proton). 
The most popular framework is provided by the conventional (so-called collinear) 
QCD factorization. In this approach, gluon and quark densities depend only on the 
longitudinal momentum fraction $x$ of the proton momentum carried by a parton. An
appropriate QCD evolution describing the dependence of PDFs on the resolution scale $\mu^2$
is given by the Dokshitzer-Gribov-Lipatov-Altarelli-Parisi (DGLAP) equations\cite{DGLAP}.
Such formalism is usually successful for sufficiently inclusive processes, like 
deep-inelastic lepton-hadron scattering (DIS), if a few higher-order 
terms in perturbative QCD expansion are taken into account.

However, in order to describe less inclusive processes proceeding at high energies
with large momentum transfer and/or containing multiple hard scales the 
Transverse Momentum Dependent (TMD, or unintegrated) parton densities $f_a(x, {\mathbf k}_T^2, \mu^2)$
with $a = q$ or $g$
are required (for more information see, for example, review\cite{TMD-review} and references therein).
These quantities encode additional transverse momentum and polarization degrees of 
freedom and satisfy the Balitsky-Fadin-Kuraev-Lipatov (BFKL)\cite{BFKL} or 
Catani-Ciafaloni-Fiorani-Marchesini (CCFM)\cite{CCFM} evolution equations.
In this way one can effectively resum large logarithmic terms proportional to
$\alpha_s^n \ln^n s/\Lambda_{\rm QCD}^2 \sim \alpha_s^n \ln^n 1/x$ which 
are expected to become equally (or even more) important in comparison with 
conventional DGLAP contributions proportional to $\alpha_s^n \ln^n \mu^2/\Lambda_{\rm QCD}^2$.
Such high-energy factorization\cite{HighEnergyFactorization}, or $k_T$-factorization\cite{kt-factorization}
formalism was formulated and it is becoming a widely exploited tool in high energy physics.
A certain advantage of this approach is that one can quite easily take into account 
a large piece of higher-order pQCD corrections into the calculations. 
Several Monte-Carlo event generators based on the $k_T$-factorization formalism, 
like as \textsc{cascade}\cite{CASCADE}, \textsc{katie}\cite{KATIE} and \textsc{pegasus}\cite{PEGASUS} are developed
and a number of corresponding phenomenological applications is known in the literature. Thus, 
the $k_T$-factorization approach becomes an essential tool which allows 
one to make theoretical predictions for future experiments at modern (LHC) and future (FCC, EiC, EicC)
colliders.

In this sence, a special interest is connected with the selection of the TMD parton density in the proton
best suited to describe the currently available collider data
and which, therefore, can be used to generate the necessary realistic predictions. 
However, in contrast with a great amount of our knowledge about the conventional PDFs 
accumulated in theoretical and experimental studies over past years,
the TMD parton densities are still poorly known quantities.
There are some popular approaches to evaluate the latter, 
for example, the Kimber-Martin-Ryskin (KMR) prescription\cite{KMR-LO, KMR-NLO}, CCFM-based formalism\cite{JH2013}
and Parton Branching (PB) approach\cite{PB1, PB2}.
Variety of currently available TMD sets
are collected in the \textsc{tmdlib} package\cite{TMDLib2}, which is a C$++$ library 
providing a framework and an interface to the different parameterizations.

It is known that an important role in derivation 
of TMD parton densities in a proton plays the appropriate choice of the non-perturbative input $f_a^{(0)}(x, {\mathbf k}_T^2, \mu_0^2)$, 
which is used as the initial condition for 
subsequent QCD evolution\cite{InputImportance-1, InputImportance-2, InputImportance-3, InputImportance-4}.
In fact, its influence on the description of experimental data can be significant\cite{Input-1, Input-2, Input-3, Input-4}.
Similar to collinear PDFs,
starting TMD parton distributions
are usually parameterized in a rather general form (see Section~2 below) and then fitted to some experimental data.
Such procedures were carried out for the CCFM\cite{JH2013}
and PB\cite{PB} approaches with \textsc{xFitter} tool\cite{xFitter},
where latest precision HERA measurements of the proton structure function $F_2(x,Q^2)$
were used.
In contrast, in our previous studies\cite{Input-1, Input-2, Input-3, Input-4}
the modified soft quark-gluon string model (QGSM)\cite{SoftQuarkGluonStringModel-1, SoftQuarkGluonStringModel-2} 
was applied to determine parameters of an analytical expression for the starting TMD gluon density
in a proton, $f_g^{(0)}(x, {\mathbf k}_T^2, \mu_0^2)$. 
In the modified QGSM both the longitudinal and transverse motion of quarks and 
gluons\cite{ModifiedSoftQuarkGluonStringModel-1, ModifiedSoftQuarkGluonStringModel-2} as well 
as the saturation effects at small $x$ and low scales can be taken into account.
The essential phenomenological parameters were obtained from the best description of 
RHIC and LHC data on the inclusive spectra of hadrons 
produced in $pp$ and $AA$ collisions at low transverse momenta,
and the CCFM evolution equation was applied
to extend the proposed TMD gluon density in the whole kinematical region.
It was shown that such an approach
is able to
describe HERA data on proton structure functions
$F_2^c(x,Q^2)$, $F_2^b(x,Q^2)$ and $F_L(x,Q^2)$
and LHC data on several processes, in particular,
single top production and inclusive Higgs boson production at $\sqrt s = 8$ and $13$~TeV.

In the present paper we continue our study 
and recalculate $f_g^{(0)}(x, {\mathbf k}_T^2, \mu_0^2)$ more accurately
using the 
modified QGSM. 
Moreover, we determine the parameters
of the initial gluon density using the LHC data on 
soft hadron (kaon and pion), inclusive $b$-jet 
and Higgs boson production in $pp$ collisions at different energies as well as
latest HERA data on 
proton structure functions $F_2^c(x, Q^2)$ and $F_2^b(x, Q^2)$
and reduced cross sections $\sigma_{\rm red}^{c}(x,Q^2)$ and $\sigma_{\rm red}^{b}(x,Q^2)$.
Thus, we perform a step forward to the global fit of TMD gluon density from the collider data,
that significantly improves our earlier analyses\cite{Input-1, Input-2, Input-3, Input-4}.

The paper is organized as follows. In Section~2 we shortly describe 
our theoretical input and discuss the basic steps of calculations of 
soft hadron spectra in the modified QGSM.
Then, from best description of the LHC data on the latter we derive an updated 
analytical expression for the initial TMD gluon density 
in a proton. In Section~3 we perform a fit of several phenomenological parameters 
from the LHC and HERA data and compare our results with the known ones.
We give conclusions in Section~4.

\section{The model} \indent

Similar to conventional PDFs,
a construction of the TMD parton distributions in a proton starts from the input densities,
which are further used as the initial conditions for subsequent non-collinear QCD evolution.
As it was mentioned above, usually the initial TMD gluon density at some 
starting scale $\mu_0^2$ (which is of order of hadron scale) is taken in the rather general empirical form 
with factorized Gauss smearing in transverse momentum ${\mathbf k}_T^2$ (see, for example,\cite{JH2013}):
\begin{gather}
  f_g^{(0)}(x, {\mathbf k}_T^2, \mu_0^2) = a_1 x^{a_2} (1-x)^{a_3} e^{-{\mathbf k}_T^2/q_0^2},
  \label{eq:Jung-Hautmann}
\end{gather}
\noindent
where all the parameters have to be extracted from the experimental data.
Alternatively, a more physically motivated non-factorized expression 
for $f_g^{(0)}(x, {\mathbf k}_T^2, \mu_0^2)$ can be taken 
from the Golec-Biernat-W\"usthoff (GBW) approach \cite{GBW1, GBW2}
based on color dipole picture for deep inelastic scattering (DIS):
\begin{gather}
  f_g^{(0)}(x, {\mathbf k}_T^2, \mu_0^2) = c_g R_0^2(x) {\mathbf k}_T^2 e^{ -R_0^2(x){\mathbf k}_T^2 }, \quad R_0(x) = {1\over Q_0} \left( {x\over x_0} \right)^{\lambda/2},
  \label{eq:GBW}
\end{gather}
\noindent
where $c_g = 3 \sigma_0/(4 \pi^2 \alpha_s)$, $\sigma_0 = 29.12$~mb, $\alpha_s = 0.2$, $Q_0 = 1$~GeV, $x_0 = 4.1 \cdot 10^{-5}$ and $\lambda = 0.277$.
In this approach, the effect of saturation of $q \bar q$ dipole cross section 
at large distance $r$ between quark and anti-quark in the dipole or small $\mu$ is assumed.
This saturation of the  dipole cross-section is a direct consequence of the saturation of the cross-section of 
virtual photon-proton scattering ($\gamma^* p$)~\cite{GBW1}. It leads to scale-independent 
behavior of the TMD gluon
density $f_g^{(0)}(x, {\mathbf k}_T^2, \mu_0^2)$ at $\mu < \mu_{\rm sat}$, where $\mu_{\rm sat}$ 
is the saturation scale. 
The GBW model was successfully applied to both inclusive and diffractive $ep$ scattering at HERA.
However, it meets some difficulties in accurate description of several hard LHC processes. 
In our previous studies\cite{Input-1, Input-2, Input-3}
to describe successfully LHC data on soft 
hadron production in $pp$ collisions 
we modified 
the starting form of the gluon density~(\ref{eq:GBW}) as the following:
\begin{equation}
  \displaystyle f_g^{(0)}(x,{\mathbf k}_T^2,\mu_0^2) = c_0 c_1 (1-x)^{b} \times \atop {
  \displaystyle \times \left[R_0^2(x){\mathbf k}_T^2 + c_2\left(R_0^2(x){\mathbf k}_T^2\right)^{a/2}\right] \exp\left(-R_0(x)
  |{\mathbf k}_T|-d\left[R_0^2(x){\mathbf k}_T^2\right]^{3/2}\right)}.
\label{def:GLLZ}
\end{equation}
\noindent
Then, to extend the consideration to a 
region of larger $p_T$
we added to~(\ref{def:GLLZ})
some function dependent on $k_T$ and low $x$, 
which was matched with the exact solution\cite{Kovchegov2000} of the BFKL equation outside 
of the saturation region. 
As it was mentioned above, 
in this way one could achieve reasonably well description\cite{Input-4} of some
HERA and LHC data using the proposed analytical expression (non-factorized with respect to $x$ and 
${\mathbf k}_T^2$) for the initial TMD gluon density with parameters obtained in the modified QGSM 
approach\cite{ModifiedSoftQuarkGluonStringModel-1, ModifiedSoftQuarkGluonStringModel-2}. 
However, in the present paper we suggest a new approach for calculation of $p_T$ spectra of soft hadron production. 
Below
we discuss our suggestion based on the scale independent behavior of the starting gluon density 
$f_g^{(0)}(x, {\mathbf k}_T^2, \mu_0^2)$ at $\mu\leq\mu_0$ and low $x$. After that
we found corresponding phenomenological parameters 
from the best description
of soft hadron spectra measured at different LHC energies. 
For the reader's convenience, below
we recall shortly the basic formulas with a brief review of calculation steps.

\subsection{Hadron spectra at low $p_T$ in the mid-rapidity region } \indent

As is well known, the soft hadron production in $pp$ collisions at small momentum transfer 
and large Feynman variables $x_F$
can be analyzed successfully  within the soft QCD models, such as quark-gluon string
model (QGSM)~\cite{SoftQuarkGluonStringModel-1, SoftQuarkGluonStringModel-2} 
or dual parton model (DPM)~\cite{DPM}. It is based on the Regge behaviour of the cross section at 
large $x_F$. 
In the QGSM, the interaction dynamics is based on
two colorless strings formed between the quark/diquark ($q/qq$) and 
diquark/quark ($qq/q$) of the colliding protons\footnote{These two colorless strings can be stretched between valence quarks and diquarks 
corresponding to the one-Pomeron exchange between colliding protons.
Also many strings can be stretched between sea quarks and antiquarks in the interacting protons, which corresponds to $n$-Pomeron exchanges}.
At their breaking, the quark-antiquark and diquark-antidiquark pairs are
created in the chromostatic QCD field and then they fragmentate into final hadrons $h$.
Corresponding quark and diquark distribution functions and their fragmentation
functions into hadrons were calculated~\cite{SoftQuarkGluonStringModel-1, SoftQuarkGluonStringModel-2}.
Such approach allows one to describe the experimental observables at non zero $x_F$ and low  transverse mpomenta $p_T$
quite satisfactorily. 

In the mid-rapidity region, according to the Abramovsky-Gribov-Kancheli cutting rules (AGK) \cite{AGKRules},
only one-pomeron Mueller-Kancheli diagrams contribute to the inclusive hadron spectrum.
However, it has some difficulties in the description of
inclusive hadron spectra measured in this kinematical region. 
In fact, the predicted hadron
transverse momentum distributions fall down very fast with increase of $p_T$ compared to the data \cite{ModifiedSoftQuarkGluonStringModel-2}. 
To avoid these difficulties the QGSM was 
modified \cite{ModifiedSoftQuarkGluonStringModel-1, ModifiedSoftQuarkGluonStringModel-2}. So, it was
suggested that there are soft gluons in a proton which split 
into $q \bar q$ pairs and, therefore, give additional contribution to the hadron spectrum.
The contribution of the one-Pomeron exchange graph between gluons in the colliding protons and contribution of one-Pomeron
Mueller-Kancheli diagrams to inclusive $p_T$ 
spectrum  were taken into account.  However, the application of the AGK cutting rules 
in the case, when soft gluons as well as quarks are included in the calculation is very questionable.
In \cite{ModifiedSoftQuarkGluonStringModel-1, ModifiedSoftQuarkGluonStringModel-2} 
these contributions of quarks, diquarks and gluons 
were calculated separately, independently of each other assuming that the contribution of gluons to the spectrum vanishes at zero 
transverse momenta of produced hadrons.  Additionally to that the splitting function of gluons into $q{\bar q}$ pairs was calculated
ignoring the dependence of gluon distribution on the transverse momentum $k_T$. 
Therefore, in this paper we recalculate the inclusive 
$p_T$ spectra of charge hadrons produced in $pp$ collisions at mid-rapidity and small $p_T$.

As mentioned above, the data~\cite{ZEUS1,ZEUS2} on $\gamma^* p$ cross-section show its  saturation at low $Q^2$ and low $x$
\cite{GBW1}. It leads to the saturation of the dipole cross section and scale 
independent behavior of the starting gluon density at low $Q^2$ less than
the saturation transfer square $Q_{\rm sat}^2$.  Therefore, the colliding protons at low $Q^2$ can be considered as two systems consisting of three
valence quarks and gluon environment with the wave function $\Psi_g$, its square is related to the starting gluon distribution as 
$|\Psi_g|^2\sim {\tilde f}_g^{(0)}(x, {\mathbf k}_T^2, \mu_0^2)$.    
Then, the $pp$ interaction amplitude can be presented in the simple spectator form
 $F_{pp} = f^{(0)}_{3q}\Psi_g$, where $f^{(0)}_{3q}$ is the amplitude of interaction of two $3q$ systems.
To calculate the inclusive spectrum 
$\rho(x,p_T)\equiv E {d^3\sigma \over d^3 {\bf p}} \equiv {1\over \pi} {d^3 \sigma \over d^2 p_T dy}$
of hadrons $h$ we have to calculate the sum of the quark contribution $\rho_q$  and the gluon one $\rho_g$, i.e.,  
\begin{gather}
 \rho(x,p_T)~= 
~\rho_q(x,p_T)~+~\rho_g(x,p_T).
\label{def:rho_qg}
\end{gather}
\noindent
The first term in (\ref{def:rho_qg}) was calculated within the QGSM 
\cite{SoftQuarkGluonStringModel-1, SoftQuarkGluonStringModel-2}
using only the one-Pomeron exchange or the cylinder graph because in the mid-rapidity
and small $x_T=2p_T/\sqrt{s}$ the multi-Pomeron exchanges result in the negligibly small contributions,
as it was shown in \cite{ModifiedSoftQuarkGluonStringModel-2}.
It is presented in the following form:
\begin{gather}
\rho_q(s,x,p_T)=\sigma_1 \phi_q(s,x,p_T),
\label{def:rho_q}
\end{gather}
where $\sigma_1$ is the cross section of the one-Pomeron exchange, see \cite{SoftQuarkGluonStringModel-1}
and references therin:
\begin{gather}
  \sigma_1 = { \sigma_P \over z}(1-e^{-z}), \quad \sigma_P=8\pi\gamma_P(s/s_0)^\Delta, \nonumber \\
  z=\frac{2C\gamma_P(s/s_0)^\Delta}{R^2+\alpha^\prime_P\ln s/s_0},.
\label{eq:sigma1}
\end{gather}
\noindent
All the parameters in~(\ref{eq:sigma1}) are found~\cite{Sinegovsky:2018vju} from experimental data on the 
total and differential cross sections of elastic $pp$ and $p\bar p$ scattering at high energies:
$\gamma_{P} = 1.27$~GeV$^{-2}$, $\Delta = 0.156$, $C = 1.8$, $R^{2} = 4.0$~GeV$^{-2}$, $\alpha^{\prime}_{P} = 0.25$~GeV$^{-2}$.
The function $\phi_q(s,x,p_T)$ is calculated within the QGSM and presented in Eq.~(\ref{eq:HadronSpectrum}).
The second one $\rho_g(x,p_T)$  is the convolution of the modified gluon distribution 
$F_g(x,k_T)$ with the fragmentation function of gluons to hadrons $D_{g\rightarrow h}$  multiplied by the integral from $|f^{(0)}_{3q}|^2$ over the intrinsic 
phase spase, which results in approximately the inelastic $pp$ cross section $\sigma_\text{in}^{pp}$, 
because the LHC data described in this paper exclude the elastic $pp$ collisions. The modifiedf gluon distribution $F_g(x,k_T)$ as well as the modified
quark and diquark distributions $F_q(x,k_T),F_{qq}(x,k_T)$ are calculated taking into account the energy-momentum conservation low, see Eqs.~(\ref{eq:Fq}-\ref{eq:Fg})
\begin{gather}
\rho_g(x,p_T)=F_g\otimes D_{g\rightarrow h}\times\sigma_\text{in}^{pp}.
\label{def:rho_g}
\end{gather}

Finally the hadron spectrum at low $x$ and low $p_T$ can be presented as the following:
\begin{gather}
 \rho(x,p_T)~=~\sigma_1 \phi_q(s,x,p_T) + \sigma_\text{in} \phi_g(s,x,p_T).
  \label{eq:HadronSpectrum}
\end{gather}
\noindent

First and second terms in~(\ref{eq:HadronSpectrum}) represent the quark/diquark and gluon contributions, respectively:
\begin{gather}
  \phi_q (s,x,p_T) = \Big\{\Phi_{q}(x_{+},p_T)\Phi_{qq}(x_{-},p_T) + \Phi_{qq}(x_{+},p_T)\Phi_{q}(x_{-},p_T)\Big\},
\label{eq:HadronSpectrumStep1a}
\end{gather}
\begin{gather}
  \phi_g (s,x,p_T) = \Big\{\Phi_g(x_+,p_T)+\Phi_g(x_-,p_T)\Big\},
  \label{eq:HadronSpectrumStep1b}
\end{gather}
\noindent
where 
\begin{gather}
  x_{\pm} = {1\over 2} \left( \pm x + \sqrt{x^2 + 4(m^2_h + p_T^2)/s} \right), \quad x = 2 \sqrt{m_h^2 + p_T^2 \over s} \sinh y,
  \label{eq:Kinematics1}
\end{gather}
\noindent
$m_h$ is the produced hadron mass and phenomenological parameters $C_q$ and $C_g$ have to be 
determined from the data.
Keeping in mind that the
proton consists of two up and one down quarks
and taking into account the non-zero parton transverse momenta,
the contributions of quark, diquark and gluon fragmentations within the modified QGSM
approach are calculated as convolutions\cite{ModifiedSoftQuarkGluonStringModel-1, ModifiedSoftQuarkGluonStringModel-2}:
\begin{gather}
  \Phi_q(x, p_T) = \int\limits_x^1 d \xi \int\limits_0^{\infty} d {\mathbf k}_T^2 \int\limits_0^{2\pi} d\phi \times \nonumber \\
  \times \left[ {2\over 3} F_u(\xi, {\mathbf k}_T^2) G_{u \to h}\left(z,|{\mathbf p}_T - z {\mathbf k}_T|\right) + {1\over 3} F_d(\xi, {\mathbf k}_T^2) G_{d \to h}\left(z,|{\mathbf p}_T - z {\mathbf k}_T|\right)\right],
  \label{eq:PhiQuark}
\end{gather}
\begin{gather}
  \Phi_{qq}(x, p_T) = \int\limits_x^1 d \xi \int\limits_0^{\infty} d {\mathbf k}_T^2 \int\limits_0^{2\pi} d\phi \times \nonumber \\
  \times \left[ {2\over 3} F_{ud}(\xi, {\mathbf k}_T^2) G_{ud \to h}\left(z,|{\mathbf p}_T - z {\mathbf k}_T|\right) + {1\over 3} F_{uu}(\xi, {\mathbf k}_T^2) G_{uu \to h}\left(z,|{\mathbf p}_T - z {\mathbf k}_T|\right)\right],
  \label{eq:PhiDiQuark}
\end{gather}
\begin{gather}
  \Phi_{g}(x, p_T) = \int\limits_x^1 d \xi \int\limits_0^{\infty} d {\mathbf k}_T^2 \int\limits_0^{2\pi} d\phi F_g(\xi, {\mathbf k}_T^2) G_{g \to h}\left(z,|{\mathbf p}_T - z {\mathbf k}_T|\right),
  \label{eq:PhiGluon}
\end{gather}
\noindent
where ${\mathbf k}_T$ is the transverse momentum of quark, diquark and/or gluon,
$z = x/\xi$ and $\phi$ is the azimuthal angle between ${\mathbf p}_T$ and ${\mathbf k}_T$.
Quark, diquark and gluon fragmentation functions to hadrons (namely, to pions and kaons),
$G_{a \to h}(z, |{\mathbf p}_T|)$ with $a = q$, $qq$ or $g$,
were calculated in the QGSM at leading (LO) and next-to-leading (NLO) orders\cite{FFs}. Corresponding 
analytical expressions are collected in Appendix A.
Functions $F_a(x, {\mathbf k}_T^2)$ involved in~(\ref{eq:PhiQuark}) --- (\ref{eq:PhiGluon})
are related to the TMD parton distributions in a proton 
taken at some scale determined by the produced hadron transverse momentum\footnote{As it was mentioned above, this scale could 
be considered as the starting scale for subsequent QCD evolution 
due to small hadron transverse momentum, $p_T \sim 1$~GeV.}. 
The functions $F_{q}(x, {\mathbf k}_T^2)$, $F_{qq}(x, {\mathbf k}_T^2)$ and $F_{g}(x, {\mathbf k}_T^2)$ 
were calculated using the energy-momentum conservation law
for quark, diquark and gluon. So, for example,
\begin{gather}
  F_q(x, {\mathbf k}_T^2) = \int\limits_{x_\pm}^1 d \xi_1 d \xi_2 \delta(1 - x - \xi_1 - \xi_2) \int d^2 {\mathbf p}_T d^2 {\mathbf q}_T \delta^{(2)}({\mathbf k}_T + 
 {\mathbf p}_T + {\mathbf q}_T) \times \nonumber \\
  \times \tilde f_q(x,{\mathbf k}_T^2) \tilde f_{qq}(\xi_1),{\mathbf p}_T^2) \tilde f_{g}(\xi_2, {\mathbf q}_T^2),
  \label{eq:Fq}
\end{gather}
\noindent
where $\tilde f_a(x, {\mathbf k}_T^2) \equiv f_a(x) g_a({\mathbf k}_T^2)$ and
$\tilde f_{g}(x, {\mathbf k}_T^2)\equiv f_{g}(x, {\mathbf k}_T^2)/x$ are the $k_T$-dependent quark, diquark and gluon densities, respectively 
(see below).
Performing integration over $\xi_1$ and ${\mathbf p}_T^2$ in~(\ref{eq:Fq}), one can easily obtain:
\begin{gather}
  F_u(x, {\mathbf k}_T^2) = \tilde f_u(x) g_q({\mathbf k}_T^2) \int\limits_{x_\pm}^{1 - x} d\xi_2 \int\limits_0^\infty d{\mathbf q}_T^2 \int\limits_0^{2\pi} d\varphi \times \nonumber \\
  \times \tilde f_{ud}(1 - x - \xi_2) g_{qq}(|{\mathbf k}_T + {\mathbf q}_T|^2) \tilde f_{g}(\xi_2, {\mathbf q}_T^2),
  \label{eq:Fu}
\end{gather}
\begin{gather}
  F_d(x, {\mathbf k}_T^2) = \tilde f_d(x) g_q({\mathbf k}_T^2) \int\limits_{x_\pm}^{1 - x} d\xi_2 \int\limits_0^\infty d{\mathbf q}_T^2 \int\limits_0^{2\pi} d\varphi \times \nonumber \\
  \times \tilde f_{uu}(1 - x - \xi_2) g_{qq}(|{\mathbf k}_T + {\mathbf q}_T|^2) \tilde f_{g}(\xi_2, {\mathbf q}_T^2),
  \label{eq:Fd}
\end{gather}
\noindent
where $\varphi$ is the azimuthal angle between ${\mathbf k}_T$ and ${\mathbf q}_T$.
Similar to that one can derive expressions for $F_{qq}(x, {\mathbf k}_T^2)$:
\begin{gather}
  F_{ud}(x, {\mathbf k}_T^2) = \tilde f_{ud}(x) g_{qq}({\mathbf k}_T^2) \int\limits_{x_\pm}^{1 - x} d\xi_2 \int\limits_0^\infty d{\mathbf q}_T^2 \int\limits_0^{2\pi} d\varphi \times \nonumber \\
  \times \tilde f_{u}(1 - x - \xi_2) g_{qq}(|{\mathbf k}_T + {\mathbf q}_T|^2) \tilde f_{g}(\xi_2, {\mathbf q}_T^2),
  \label{eq:Fud}
\end{gather}
\begin{gather}
  F_{uu}(x, {\mathbf k}_T^2) = \tilde f_{uu}(x) g_{qq}({\mathbf k}_T^2) \int\limits_{x_\pm}^{1 - x} d\xi_2 \int\limits_0^\infty d{\mathbf q}_T^2 \int\limits_0^{2\pi} d\varphi \times \nonumber \\
  \times \tilde f_{d}(1 - x - \xi_2) g_{qq}(|{\mathbf k}_T + {\mathbf q}_T|^2) \tilde f_{g}(\xi_2, {\mathbf q}_T^2).
  \label{eq:Fuu}
\end{gather}
\noindent
For $F_g(x, {\mathbf k}_T^2)$ we have the following formula including the charge and isotopic invariance:
\begin{gather}
  F_{g}(x, {\mathbf k}_T^2) = \tilde f_{g}(x, {\mathbf k}_T^2) \int\limits_{x_\pm}^{1 - x} d\xi_2 \int\limits_0^\infty d{\mathbf q}_T^2 \int\limits_0^{2\pi} d\varphi \times \nonumber \\
  \times \Bigg\{ {2\over 3} \tilde f_{u}(1 - x - \xi_2) g_{q}(|{\mathbf k}_T + {\mathbf q}_T|^2) \tilde f_{ud}(\xi_2) g_{qq}({\mathbf q}_T^2) + \nonumber \\ 
  + {1\over 3} \tilde f_{d}(1 - x - \xi_2) g_{q}(|{\mathbf k}_T + {\mathbf q}_T|^2) \tilde f_{uu}(\xi_2) g_{qq}({\mathbf q}_T^2) \Bigg\}.
  \label{eq:Fg}
\end{gather}
\noindent
The distributions $f_u,f_d,f_{uu},f_{ud}$ and $g_q,g_{qq}$ as functions of $x$ and $k_T$ respectively are presented in the Appendix A.
Our choice for the TMD parton densities involved in~(\ref{eq:Fu}) --- (\ref{eq:Fg}) is particularly discussed in Section~2.2.

Finally, the inelastic $pp$ cross section $\sigma_\text{in}$ can be calculated 
as the difference between the total and elastic $pp$ scattering cross sections: $\sigma_\text{in} = \sigma_\text{tot} - \sigma_\text{el}$,
where $\sigma_\text{tot}$ should satisfy to Regge asymptotic, $\sigma_\text{tot} \sim (s/s_0)^{\alpha_P - 1}$.
However, it was shown\cite{sigmas2013,Dremin2019} 
that $\sigma_\text{tot}$ and $\sigma_\text{el}$ can be parametrized in the following way:
\begin{gather}
  \sigma_\text{tot} = 21.7(s/s_0)^{0.0808} + 56.08(s/s_0)^{-0.4525} \text{ mb}, \\
  \sigma_\text{el} = 11.84-1.617\ln s + 0.1359 \ln^2 s \text{ mb}.
\label{eq:sigmaTOT}
\end{gather}
\noindent
where $s_0 = 1$~GeV. Therefore, we will use these expressions in the numerical calculations. 

Using the expressions above, one can calculate the cross sections for soft hadron production
in $pp$ collisions. Some technical details are given in Appendix B. 
To perform numerical multidimensional integration, we employ
a Monte-Carlo technique implemented into the \textsc{vegas} tool\cite{VEGAS}. 

\subsection{TMD parton distributions in a proton at low scale} \indent

Concerning the TMD gluon density, $f_g(x, {\mathbf k}_T^2)$, 
here we will follow our previous considerations\cite{Input-1, Input-2, Input-3, Input-4}
where the different expressions based on the GBW saturation model were tried.
In fact, it was demonstrated that overall description of collider data
could be significantly improved if usual GBW gluon distribution given by~(\ref{eq:GBW})
is modified. In the present analysis we update the modification
proposed earlier\cite{Input-1, Input-2} and take into account certain quark-gluon sum 
rule\footnote{The TMD gluon density used in\cite{Input-3, Input-4} has a very high ${\mathbf k}_T^2$
tail, that leads to sizeble value of gluon average transverse momentum.
The latter, of course, should have a significant perturbative component unwanted for our purposes.}: 
\begin{equation}
  \sum_a \int\limits_0^1 dx \int d^2{\mathbf k}_T \, x{\tilde f}_a(x, {\mathbf k}_T^2) = 1.
  \label{eg:SumRule}
\end{equation}
\noindent
where $a = u$, $d$, $uu$, $ud$ and $g$.

We will consider the data on charged hadron (pion and kaon) production at 
small transverse momenta $p_T \leq 1$~GeV
taken at different energies, namely, $\sqrt s = 0.9$, $2.36$, $7$ and $13$~TeV~\cite{ATLAS-soft,CMS-soft,SH-13}. 
We find that in order to describe these data the most appropriate expression for the starting gluon density 
$f_g(x, {\mathbf k}_T^2.\mu_0^2)$, hereafter labeled as LLM gluon, is the following:
\begin{equation}
  f_g(x, {\mathbf k}_T^2) = c_g (1-x)^{b_g} \sum_{n = 1}^3 \left(c_n R_0(x) |{\mathbf k}_{T}|\right)^n e^{-R_0(x)|{\mathbf k}_{T}|},
  \label{eg:OurGluon}
\end{equation}
\noindent
where $R_0(x)$ is defined in~(\ref{eq:GBW}) and we kept $x_0 = 4.1 \cdot 10^{-5}$
and $\lambda = 0.22$.
Our best fit for phenomenological parameters
leads to $c_1 = 5$, $c_2 = 3$, $c_3 = 2$ and $Q_0 = 1.233$~GeV.
However, the measured hadron spectra~\cite{ATLAS-soft,CMS-soft,SH-13} refer to relatively small $x$ region and 
appear to be mostly insensitive to  
$b_g$ value, which plays a role at essentially larger $x$.
So, we determined the latter from the LHC 
data on several hard processes, as described in Section~3.
Following \cite{KMR-DAS-1, KMR-DAS-2}, we treat $b_g$ as a running parameter at ${\mathbf k}_T^2 \geq Q_0^2$: 
\begin{equation}
  b_g = b_g(0) + {4 C_A \over \beta_0} \ln {\alpha_s(Q_0^2) \over \alpha_s({\mathbf k}_T^2)},
  \label{eg:bg}
\end{equation}
\noindent
where $b_g(0) = 5.854$, $C_A = N_c$ and 
$\beta_0 = 11 - 2/3 N_f$ is the first coefficient of the QCD $\beta$-function.
At small ${\mathbf k}_T^2 < Q_0^2$, the fixed value $b_g = b_g(0)$ is used.
A similar approach was applied in the investigation of EMC effect from 
the study of shadowing at low $x$ to antishadowing at $x \sim 0.1 - 0.2$\cite{bgForm}.

\begin{figure}
\begin{center}
\includegraphics[width=7.9cm]{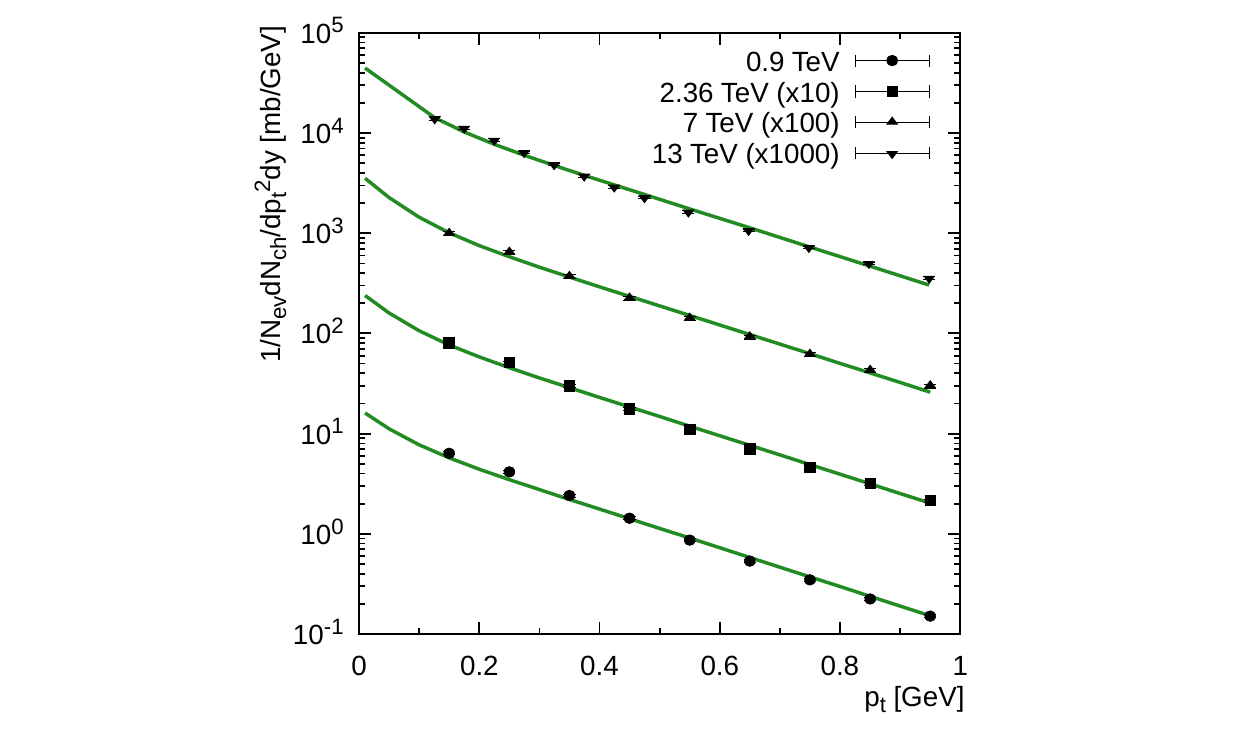}
\includegraphics[width=7.9cm]{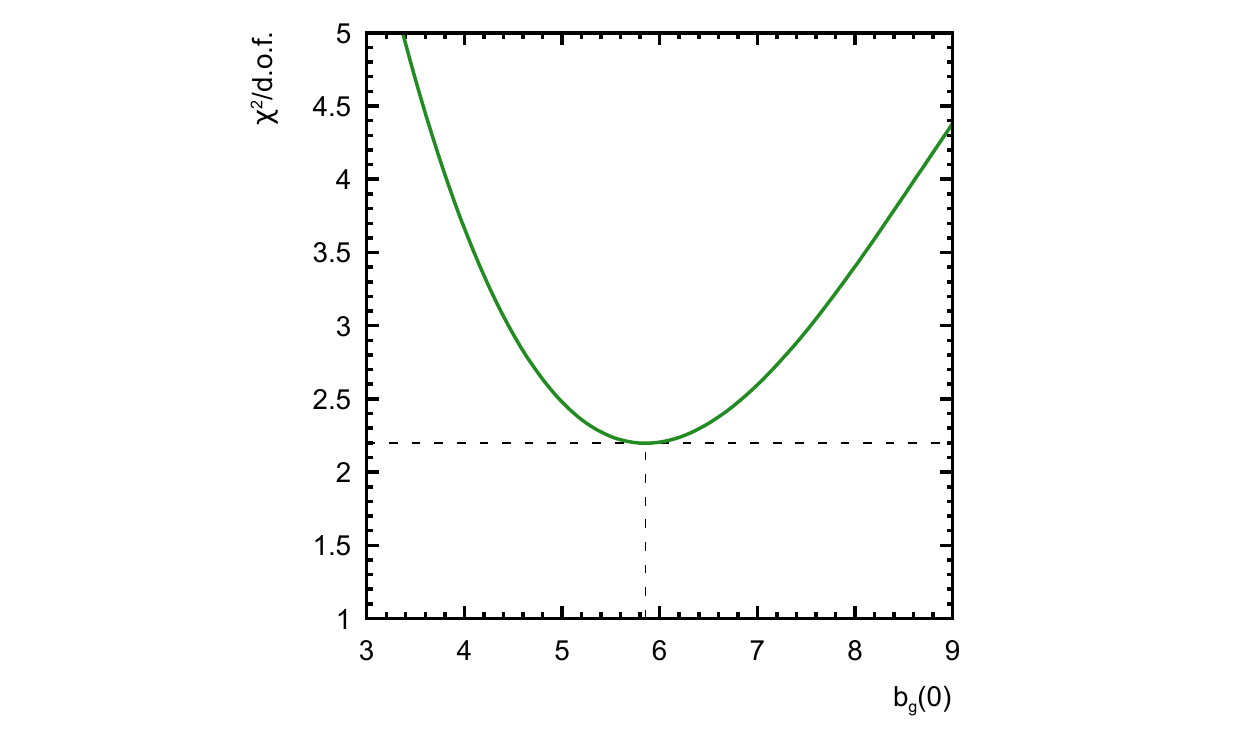}
\caption{Left panel: charged hadron spectra calculated within the 
modified QGSM at different energies. Experimental data are from~\cite{ATLAS-soft,CMS-soft,SH-13}.
Right panel: $\chi^2/d.o.f.$ dependence of our fit performed moderate 
and large $x$ values.}
\label{fig:chi2-bg0}
\end{center}
\end{figure}



The experimental data on charged hadron production involved into
the fit are compared with our predictions in Fig.~\ref{fig:chi2-bg0} (left panel). 
One can see that good agreement is achieved in a wide range of energies.

\subsection{Saturation dynamics} \indent

As it is assumed in~\cite{GBW1, GBW2}, the efffective dipole cross-section, as a function
of the distance $r$ between $q$ and $\bar q$ is saturated at large $r$. It is
presented in the following form:
\begin{equation}
\hat{\sigma}(x,r^2)~=~\sigma_0\left\{1-\exp\left(-\frac{r^2}{4R_0^2(x)}\right)\right\},\
\label{def:GBW_crsect}
\end{equation}
where $R_0$ is  determined in (\ref{eq:GBW}).
The normalization $\sigma_0$, the parameters $\lambda$ and $x_0$ were found from
a fit of all inclusive DIS data\cite{GBW1, GBW2}. The relation of the TMD
gluon distribution to the dipole cross section
$\hat{\sigma}(x,r^2)$ was calculated\cite{GBW2} within the two gluon
exchange approximation between $q,{\bar q}$ and the nucleon debris. It has the
following form:
\begin{equation}
\hat{\sigma}(x,r^2)~=~\frac{4\pi^2}{3}\int\frac{dk_t^2}{k_t^2}\left\{1-J_0(k_Tr)\right\}
\alpha_s(\mu_0^2)f_g^{(0)}(x, {\mathbf k}_T^2, \mu_0^2)~,
\label{def:GBW_dpcrsect}
\end{equation}
 where $J_0(k_Tr)$ is the cylindrical special function of order $0$. 
Comparing (\ref{def:GBW_crsect}) to (\ref{def:GBW_dpcrsect}) one can get
immediately the expression (\ref{eq:GBW}).

\begin{figure}
\begin{center}
\includegraphics[width=7.9cm]{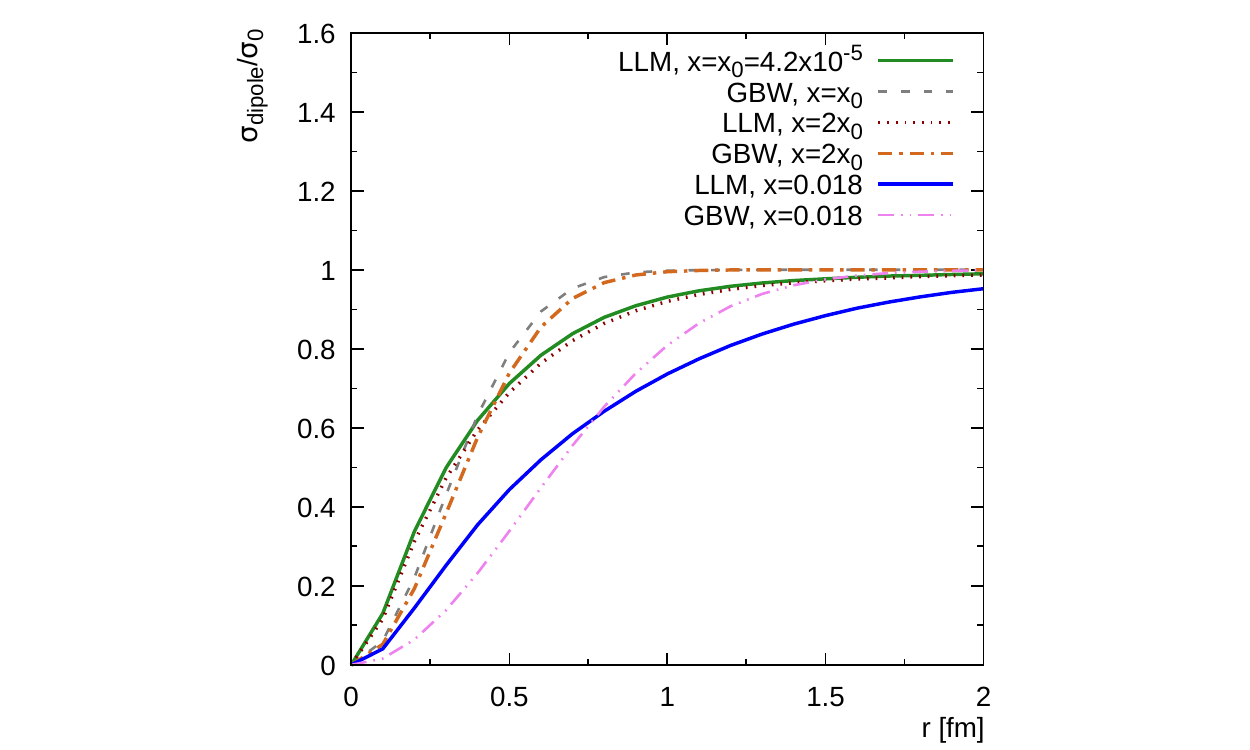}
\includegraphics[width=7.9cm]{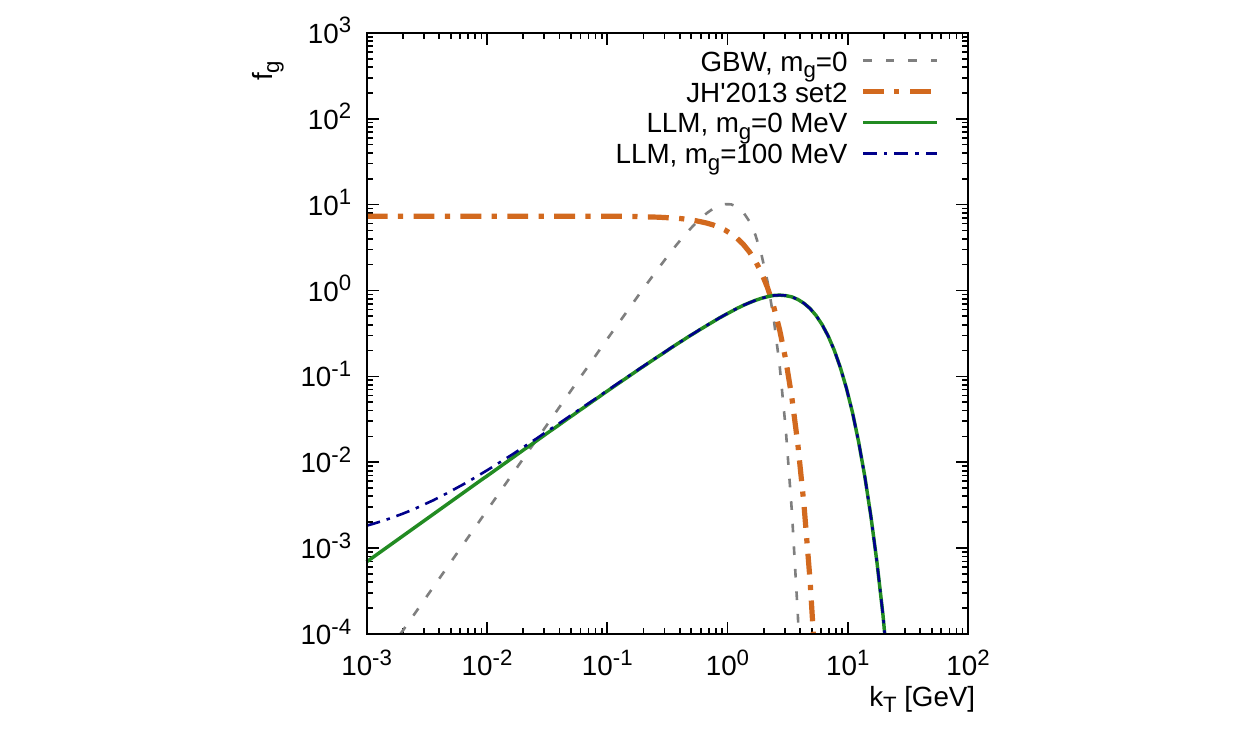}
\caption{Left: The dipole cross section as a function of $r$ at different values of
$x$. Right: The LLM and GBW gluon densities as functions of $k_T$ at different 
gluon masses. The JH'2013 set 2\cite{JH2013} gluon distribution is presented for comparison.
}
\label{fig:dipole_crsect}
\end{center}
\end{figure}

Inserting our gluon distribution (LLM) at the initial $\mu_0$ presented in (\ref{eg:OurGluon}) to
(\ref{def:GBW_dpcrsect}) one can get the dipole cross section at different values
of $x$ as a function of $r$, which is proportional to $2/Q_0$, according to
\cite{GBW2}.
In Fig.~\ref{fig:dipole_crsect} (left panel) the comparison of the GBW dipole cross section
$\hat{\sigma}(x,r^2)$ calculated using (\ref{eq:GBW}) to our calculation
(\ref{eg:OurGluon}) is presented as a functon of $r$ at different $x$. 
One can see that the saturation of the
dipole cross section at large $r$ strongly depends on $x$ and 
on the TMD gluon density. The GBW gluon 
results in the saturation scale $r_s\sim 2/R_0$, according
to (\ref{def:GBW_crsect})~\cite{GBW2}. According to
Fig.~\ref{fig:dipole_crsect},
the saturation scale corresponding to the GBW gluon density at $x=4.2\times
10^{-5}$ is
$Q_s\simeq 2/r_s\simeq 0.8$ GeV, whereas LLM gluon
results in $Q_s\simeq 2/r_s\simeq 0.4$ GeV at the same $x$. It
means that at $Q^2<Q_s^2$  the dipole cross section and starting gluon density do not depend on scale $Q^2$.

Let us note that the GBW and LLM gluon densities vanish at $k_T\rightarrow 0$. It is due to the neglection of the initial gluon mass
$m_g$ in the gluon propagator, see (\ref{def:GBW_dpcrsect}) and~\cite{GBW2}. 
With the inclusion of the gluon mass the TMD gluon 
distributions does not vanish at the zero gluon transverse momentum. 
However, $f_g^{(0)}(x, {\mathbf k}_T^2, \mu_0^2)$ does not saturate even at 
$m_g=$ 100 MeV, according to Fig.~\ref{fig:dipole_crsect} (right panel).
 
\subsection{CCFM evolution} \indent

Being sure that proposed TMD gluon density in a 
proton is able to reproduce well the collider data
in a soft kinematical region, one can 
consider the expression~(\ref{eg:OurGluon}) as the starting condition 
for subsequent QCD evolution.

As it was mentioned above, 
we will apply the CCFM equation\cite{CCFM}.
It resums both large logarithms $\alpha_s^n \ln^n 1/x$ and $\alpha_s^n \ln^n 1/(1 - x)$ 
and introduces angular ordering condition to treat correctly gluon coherence effects.
In the limit of asymptotic high energies (i.e. small $x$), it is almost equivalent to BFKL\cite{BFKL}, 
but also similar to usual DGLAP evolution\cite{DGLAP} for large $x$.
Therefore, it provides a suitable tool for our purposes.

In the leading logarthmic approximation (LLA), the CCFM equation
for TMD gluon density\footnote{Hereafter, we denote the evolution variable as $\mu^2$. 
Another notation, namely, $\bar q^2$, is also often used in the literature.} $f_g(x,{\mathbf k}_T^2,\mu^2)$
can be written as
\begin{equation}
  \displaystyle f_g(x,{\mathbf k}_T^2,\mu^2) = f_g^{(0)}(x,{\mathbf k}_T^2,\mu_0^2) \Delta_s(\mu,\mu_0) + \atop {
  \displaystyle + \int\frac{dz}{z}\int\frac{dq^2}{q^2}\Theta(\mu-zq)\Delta_s(\mu,zq) \tilde P_{gg}(z,{\mathbf k}_T^2, q^2) f_g\left(\frac{x}{z},{\mathbf k}^{\prime \, 2}_T,q^2\right) },
  \label{eq:CCFM}
\end{equation}

\noindent
where ${\mathbf k}_T^\prime = {\mathbf q}(1 - z) + {\mathbf k}_T$
and $\tilde P_{gg}(z,{\mathbf k}^2_T,q^2)$ is the CCFM
splitting function:
\begin{equation}
  \displaystyle \tilde P_{gg}(z,{\mathbf k}^2_T,q^2) = \bar\alpha_s(q^2(1-z)^2) \left[\frac{1}{1-z}+\frac{z(1-z)}{2}\right] + \atop {
  \displaystyle + \bar\alpha_s({\mathbf k}_T^2)\left[\frac{1}{z}-1+\frac{z(1-z)}{2}\right]\Delta_{ns}(z,{\mathbf k}^2_T,q^2) }.
\end{equation}

\noindent
The Sudakov and non-Sudakov (or Regge) form factors read:
\begin{equation}
 \ln \Delta_s(\mu,\mu_0)= - \int\limits_{\mu_0^2}^{\mu^2}\frac{d\mu^{\prime \, 2}}{\mu^{\prime \, 2}}\int\limits_0^{1-\mu_0/\mu^{\prime}}d\zeta\,\frac{\bar\alpha_s(\mu^{\prime \, 2}(1-\zeta)^2)}{1-\zeta},
\end{equation}
\begin{equation}
\ln \Delta_{ns}(z,{\mathbf k}_T^2, {\mathbf q}_T^2) = -\bar\alpha_s({\mathbf k}_T^2)\int\limits_0^1\frac{dz^\prime}{z^\prime}\int\frac{dq^2}{q^2}\Theta({\mathbf k}_T^2-q^2)\Theta(q^2-z^{\prime\,2} {\mathbf q}^2_T),
\end{equation}
\noindent
where $\bar \alpha_s = 3 \alpha_s/\pi$.
The first term in~(\ref{eq:CCFM}) is the initial
TMD gluon density $f_g^{(0)}(x,{\mathbf k}_T^2,\mu_0^2)$ 
determined at the scale $\mu_0^2$
multiplied by the Sudakov form factor,
describing the contribution of non-resolvable branchings between
the starting scale $\mu_0^2$ and scale $\mu^2$.
In our calculations, the expression~(\ref{eg:OurGluon}) will be used as the initial gluon density.
The second term represents the details of the QCD evolution
expressed by the convolution of the CCFM gluon splitting function $\tilde P_{gg}(z,{\mathbf k}^2_T,q^2)$
with the TMD gluon density and Sudakov form factor. The angular ordering condition
is introduced with the theta function, so
the evolution scale $\mu^2$ is coming from 
the maximum allowed angle for any gluon emission 
determined by the hard scattering subprocess: 
$\mu^2 = \hat s + {\mathbf Q}_T^2$, where
${\mathbf Q}_T$ is the net transverse momentum 
entering into the hard subprocess with center-of-mass energy $\hat s$.
This choice for scale $\mu^2$ is usually considered as a built-in property 
of the CCFM evolution (see, for example, \cite{JH2013} and references therein).

The CCFM equation can be solved numerically using 
the \textsc{updfevolv} routine\cite{uPDFevolv}, so that
the TMD gluon density can be obtained in the whole kinematical range.
In this way, the TMD gluon distribution
is tabulated in a commonly recognized format (namely, grid of $50 \times 50 \times 50$ bins in $x$, ${\mathbf k}_T^2$ and $\mu^2$)
which is used in the \textsc{tmdlib} tool\cite{TMDLib2}.

\section{Fitting the essential parameters} \indent

There are phenomenological parameters in the initial TMD gluon 
density~(\ref{eg:OurGluon}) which are not predicted by the theory
and therefore have to be extracted from the collider data.
Our fitting strategy is based on the splitting the overall procedure into two
almost independent parts, where each of them is 
referring to the regions of low and large $x$, respectively.
The low $x$ region has been already considered above, in Section~2.2.
Now we turn to moderate and large $x$ and
refine the behaviour of proposed gluon distribution
by extracting the $c_g$ and 
$b_g(0)$ parameter from measured cross sections of some hard processes.
We will use the CMS data on inclusive $b$-jet 
production in $pp$ collisions at $\sqrt s = 7$~TeV\cite{bjet-CMS-7} and 
recent data on inclusive 
Higgs boson production in different decay modes taken by the 
ATLAS\cite{HVV-ATLAS-13, HZZ-ATLAS-13} and CMS\cite{HVV-CMS-13} Collaborations at $\sqrt s = 13$~TeV.
The cross sections of these processes are governed by 
gluon-gluon fusion subprocesses and 
receive an essential contribution, in particular, from appropriate $x$ region\footnote{We will not consider 
the ATLAS data\cite{bjet-ATLAS-7} on $b$-jet production since they refer to extremely
large $b$-jet transverse momenta, where effects of parton showers and/or hadronization corrections
play an important role.}.
In fact, the CMS data on $b$-jet production refer to the kinematical region
defined by $18 < p_T(b) < 196$~GeV and rapidity $|y(b)| < 2.2$\cite{bjet-CMS-7}. 
The ATLAS data on inclusive Higgs production in diphoton decay mode were obtained at
$p_T^\gamma/m^{\gamma \gamma} > 0.35$ $(0.25)$ for the leading (subleading) decay photon, 
pseudorapidity $|\eta^\gamma| < 2.37$ for the both photons and the invariant mass $105 < m^{\gamma \gamma} < 160$~GeV\cite{HVV-ATLAS-13}.
The CMS data refer to a similar kinematical region: $p_T^\gamma/m^{\gamma \gamma} > 1/3$ $(1/4)$ for 
the leading and subleading photons, $|\eta^\gamma| < 2.5$ and $100 < m^{\gamma \gamma} < 180$~GeV\cite{HVV-CMS-13}.
The ATLAS measurement\cite{HZZ-ATLAS-13} performed in the $H \to ZZ^* \to 4l$ decay mode 
requires at least four leptons in the event with at least one lepton having
$p_T > 20$~GeV, another lepton having $p_T > 15$~GeV and 
remaining ones having $p_T > 10$ and $5$~GeV, respectively. 
All leptons must have the pseudorapidity $|\eta(l)| < 2.7$, 
the leading pair invariant mass $m_{12}$ must be $50 < m_{12} < 106$~GeV
and the subleading one should be $12 < m_{34} < 115$~GeV.
Finally, the four-lepton invariant mass $m_{4l}$ must satisfy $105 < m_{4l} < 160$~GeV cut. 
Thus, the typical $x$ values probed in these analyses, 
$x \sim 2 \mu/\sqrt s$, are varied from $x \sim 5 \cdot 10^{-3}$ to $x \sim 2 \cdot 10^{-2}$,
where the scale $\mu$ of considered processes is determined, for example, 
by the transverse masses of produced particles.
Additionally, we use the latest HERA data on the 
charm and beauty contributions to inclusive proton structure functions
$F_2^c(x, Q^2)$, $F_2^b(x, Q^2)$~\cite{F2cb-ZEUS, F2c-H1, F2cb-H1} and 
reduced cross sections
$\sigma_{\rm red}^c(x, Q^2)$ and $\sigma_{\rm red}^b(x, Q^2)$\cite{sigma_red-ZEUS+H1},
obtained at $Q^2 > 2.5$~GeV$^2$ in a wide region of $x$.
The DIS reduced cross section of heavy quark $Q$ can be presented as
\begin{gather}
  \sigma_{\rm red}^Q(x, Q^2) = F_2^Q(x, Q^2) - {y^2 \over 1 + (1 - y)^2} F_L^Q(x, Q^2),
\end{gather}
\noindent
where $y = Q^2/(xS)$ and $F_L^Q(x, Q^2)$ is the contribution
of heavy quark $Q$ to the proton longitudinal structure function $F_L(x, Q^2)$.
All these observables are governed by photon-gluon fusion
subprocess and therefore also very sensitive to the gluon content of a proton.

\begin{figure}
\begin{center}
\includegraphics[width=7.9cm]{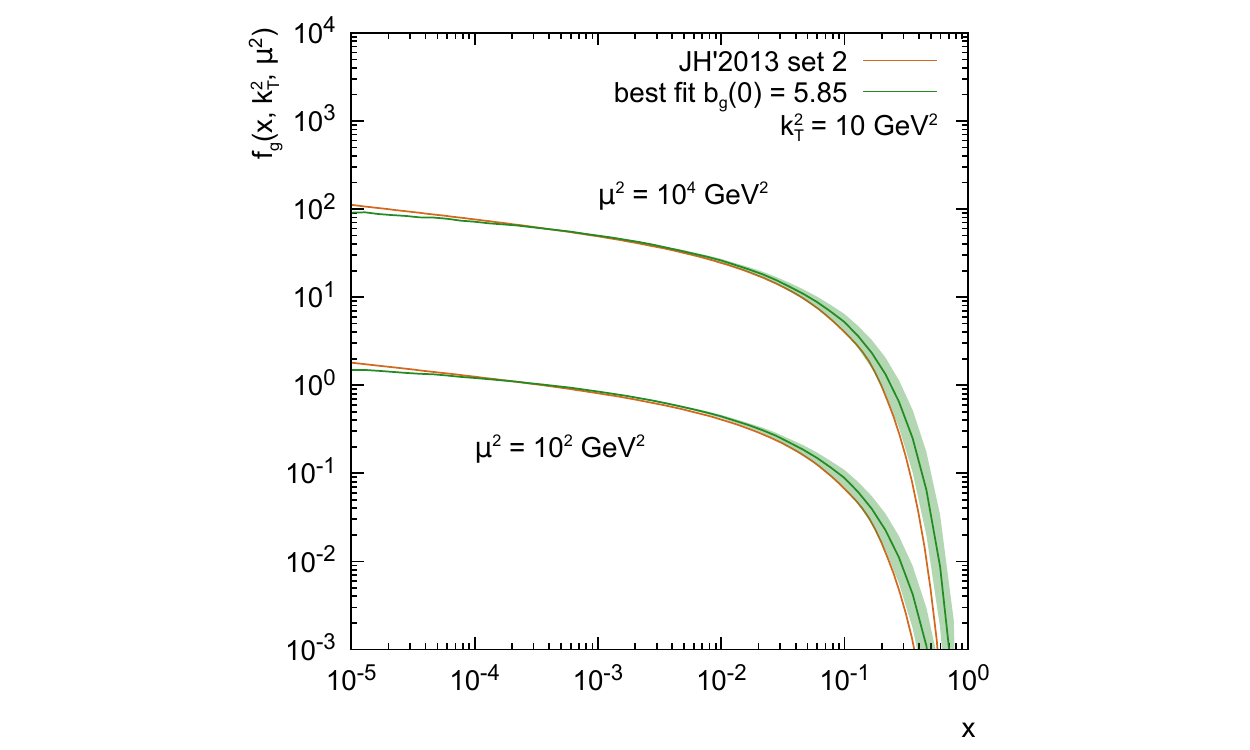}
\includegraphics[width=7.9cm]{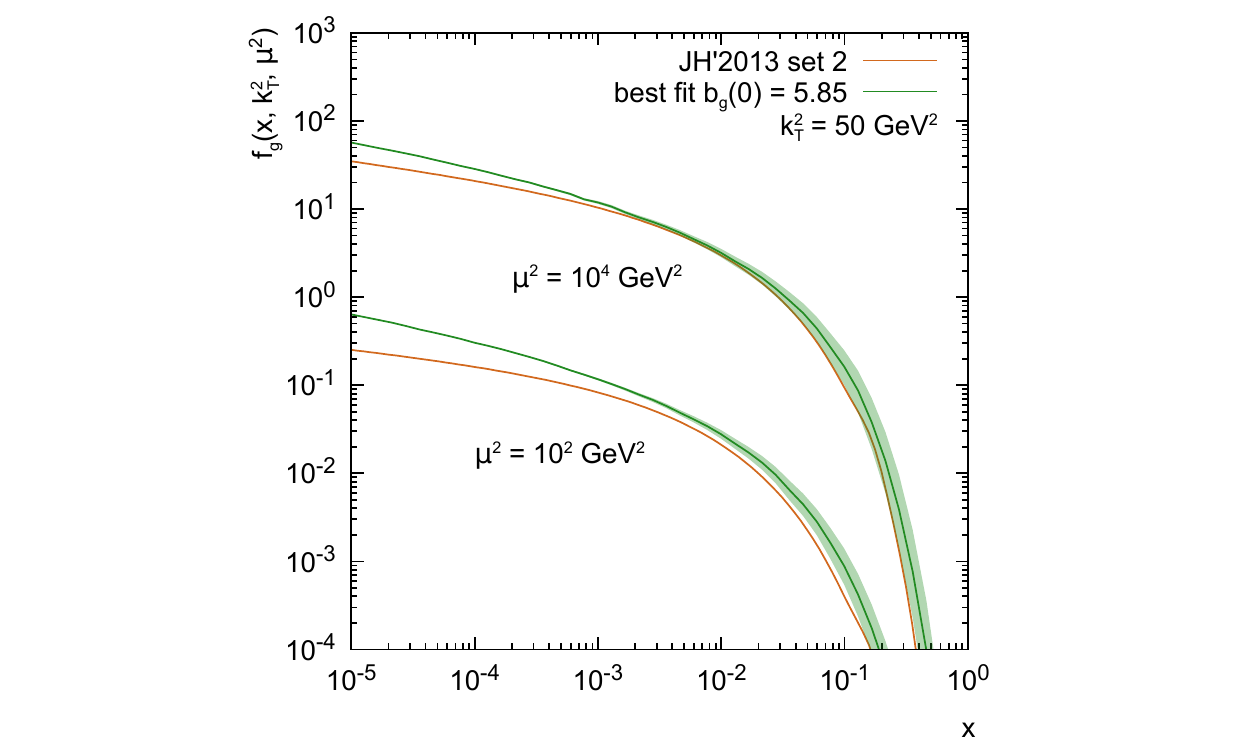}
\includegraphics[width=7.9cm]{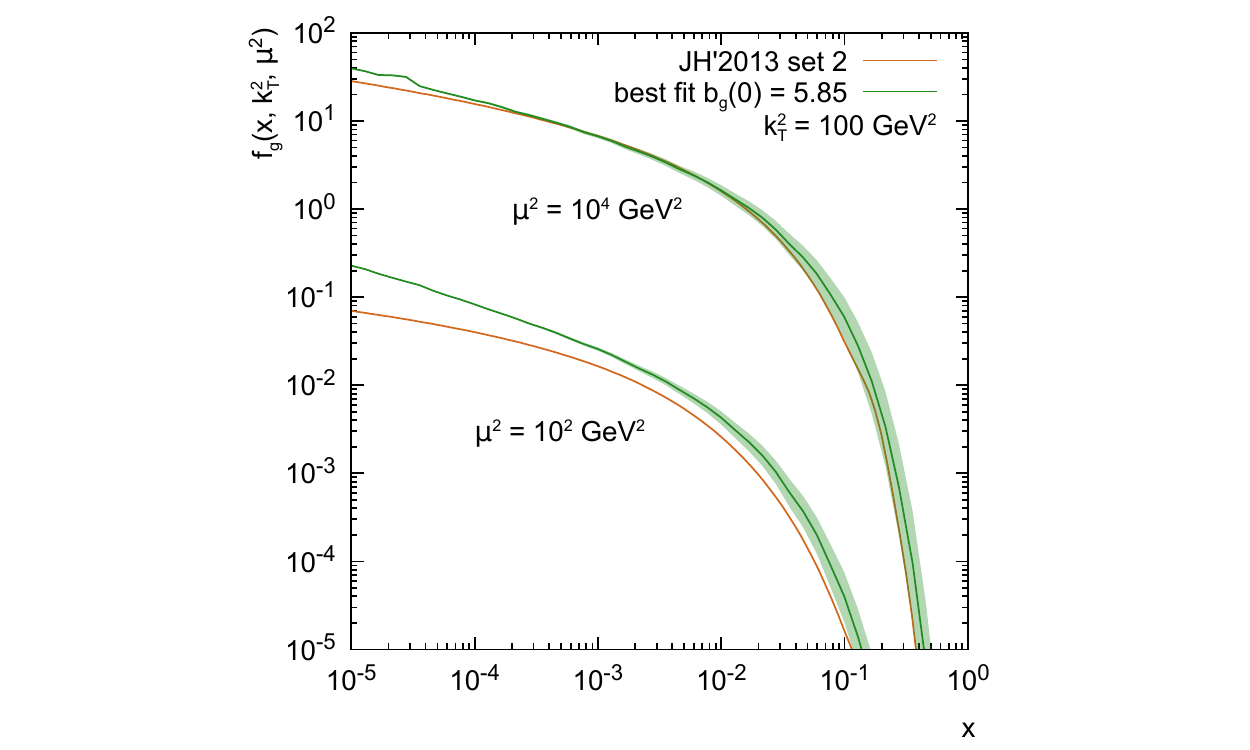}
\includegraphics[width=7.9cm]{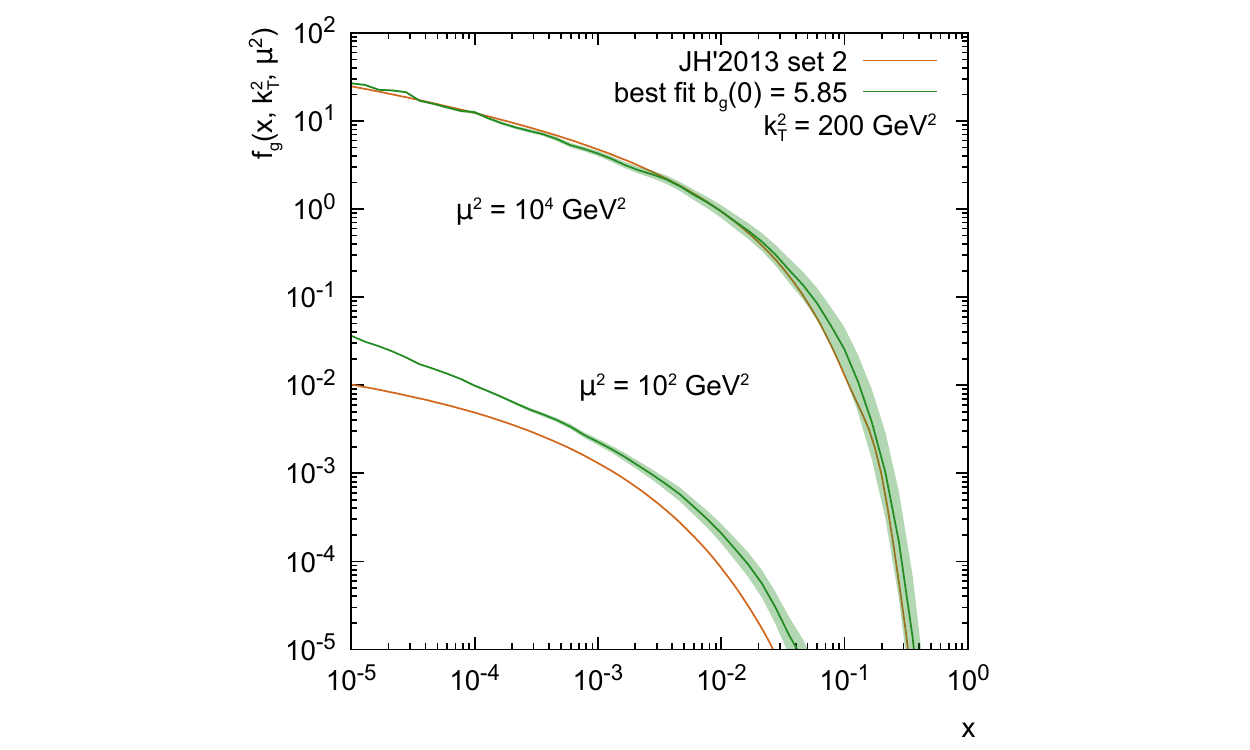}
\caption{The TMD gluon densities in a proton $f_g(x, {\mathbf k}_T^2, \mu^2)$ 
calculated as a function of longitudinal 
momentum fraction $x$ at different values of transverse momentum ${\mathbf k}_T^2$ and hard scale $\mu^2$.
Shaded bands represent the uncertainties of $b_g(0)$ fitting procedure. 
Note that the gluon densities calculated at $\mu^2 = 10^4$~GeV$^2$ are multiplied by factor of $100$.}
\label{fig:Gluonx}
\end{center}
\end{figure}

\begin{figure}
\begin{center}
\includegraphics[width=7.9cm]{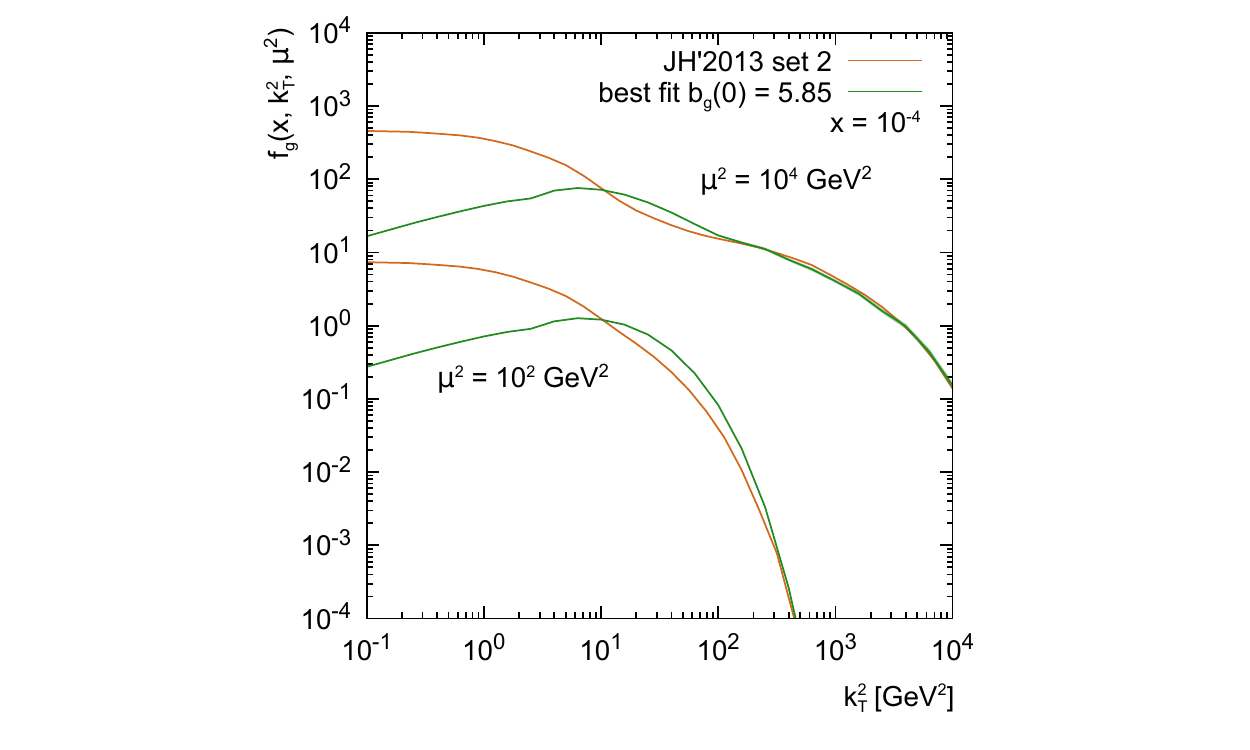}
\includegraphics[width=7.9cm]{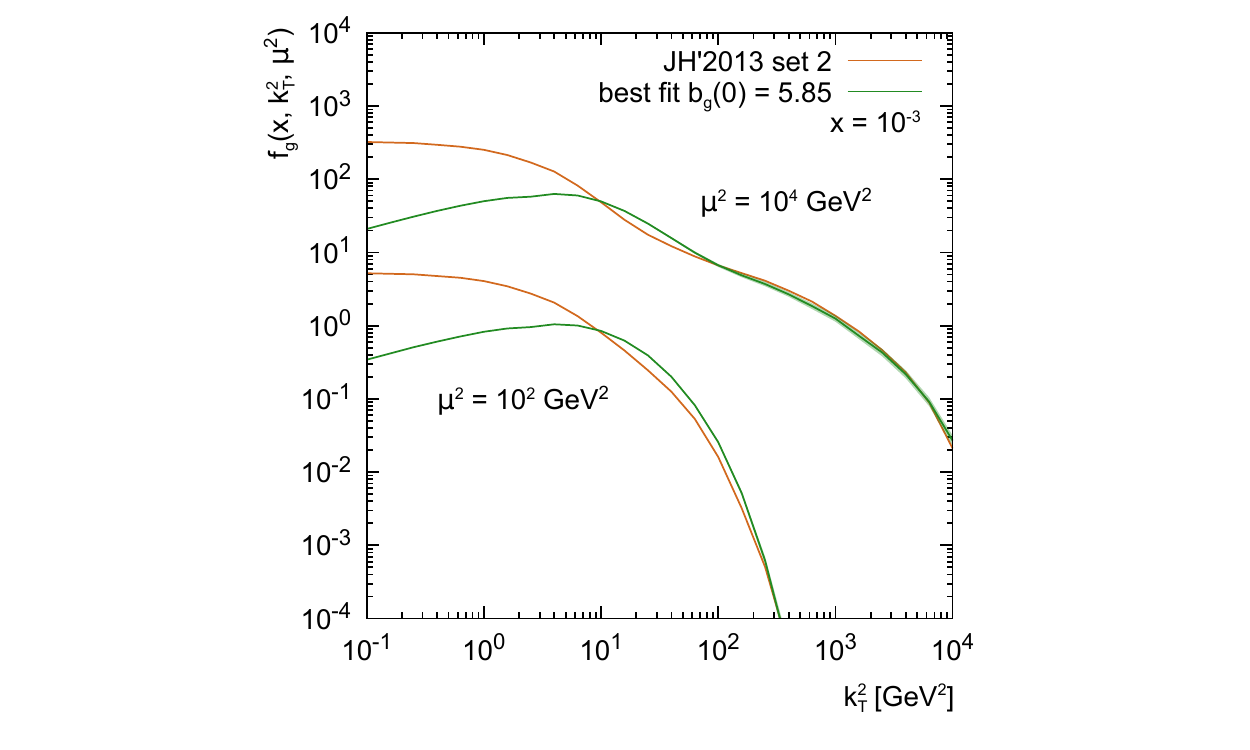}
\includegraphics[width=7.9cm]{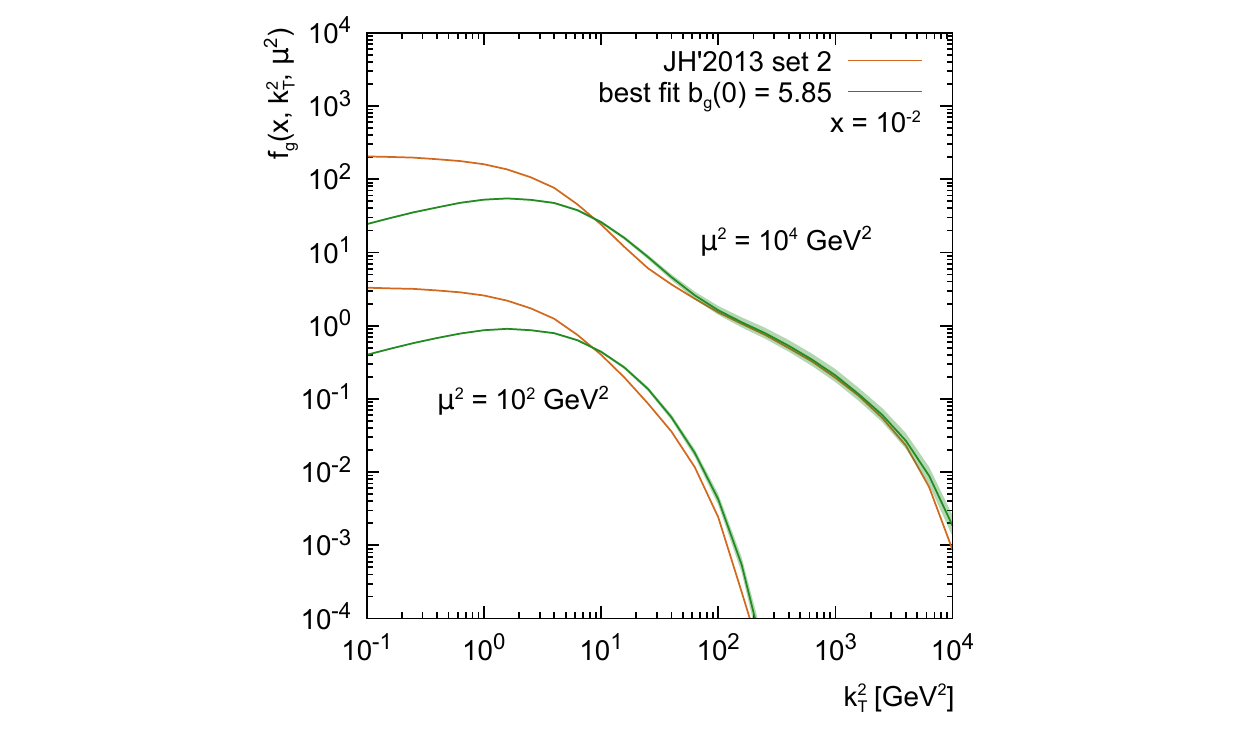}
\includegraphics[width=7.9cm]{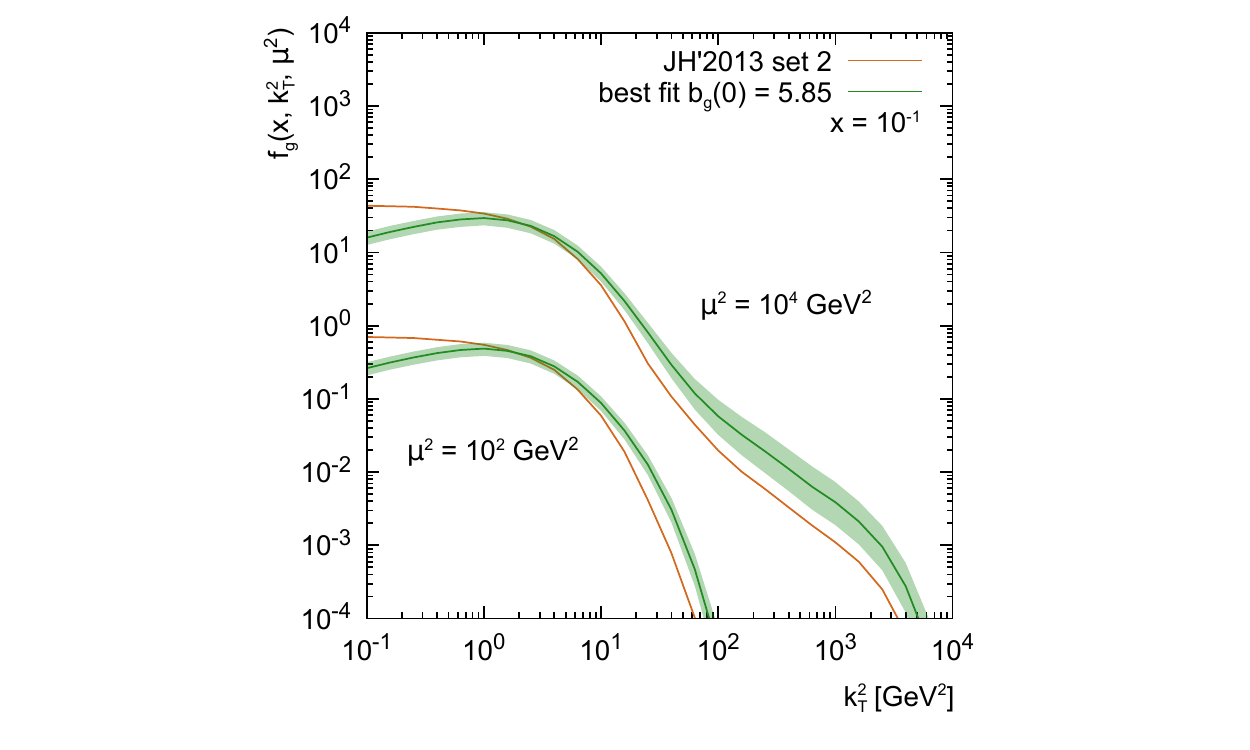}
\caption{
The TMD gluon densities in a proton $f_g(x, {\mathbf k}_T^2, \mu^2)$ 
calculated as a function of transverse momentum ${\mathbf k}_T^2$ at 
different values of longitudinal 
momentum fraction $x$ and hard scale $\mu^2$.
Shaded bands represent the uncertainties of $b_g(0)$ fitting procedure. 
Note that the gluon densities calculated at $\mu^2 = 10^4$~GeV$^2$ are multiplied by factor of $100$.}
\label{fig:Gluonkt}
\end{center}
\end{figure}

We extracted the $c_g$ and $b_g(0)$ values from best simultaneous description
of several observables, in particular, distributions on leading $b$-jet transverse momenta
measured at their different rapidities, Higgs boson transverse momentum and rapidity spectra.
We also considered 
several angular correlations in Higgs boson production, namely,
distributions on Higgs decay photon helicity angle (in the Collins-Soper frame),
leading lepton pair decay angle with respect to the beam axis (in the four-lepton rest frame)
and production angles of anti-leptons from the two decay $Z$ bosons, where these
angles are defined relatively to the $Z$ direction.
The fitting procedure is rather standard and straightforward.
Technically, applying the \textsc{updevolv} routine\cite{uPDFevolv}, we solved numerically the CCFM equation 
for a (large) number of fixed guessed $b_g(0)$ values in a wide (but still reasonable) range 
$3 < b_g(0) < 8$. Then, using each of the generated TMD gluon densities in the proton, we calculated the cross sections 
of all considered processes 
according to previous evaluations\cite{bb-kt-our, Higgs-kt-our, SFs-kt-our}. 
Best simultaneous description of the 
experimental data for all observables above is 
achieved at $c_g = 0.1731$ and 
$b_g(0) = 5.854_{-1.553}^{+1.920}$ with $\chi^2/d.o.f. = 2.2$, see Fig.~\ref{fig:chi2-bg0}.
Note that we took into account contributions 
to the Higgs production cross sections
from weak boson fusion
($W^+W^- \to H$ and $ZZ \to H$), associated $HZ$ or $HW^\pm$ production and 
associated $t \bar t H$ production. 
These contributions are essential at high transverse momenta
and have been calculated in the conventional NLO pQCD. 
We take them from\cite{HVV-ATLAS-13, HZZ-ATLAS-13, HVV-CMS-13}.

\begin{figure}
\begin{center}
\includegraphics[width=7.9cm]{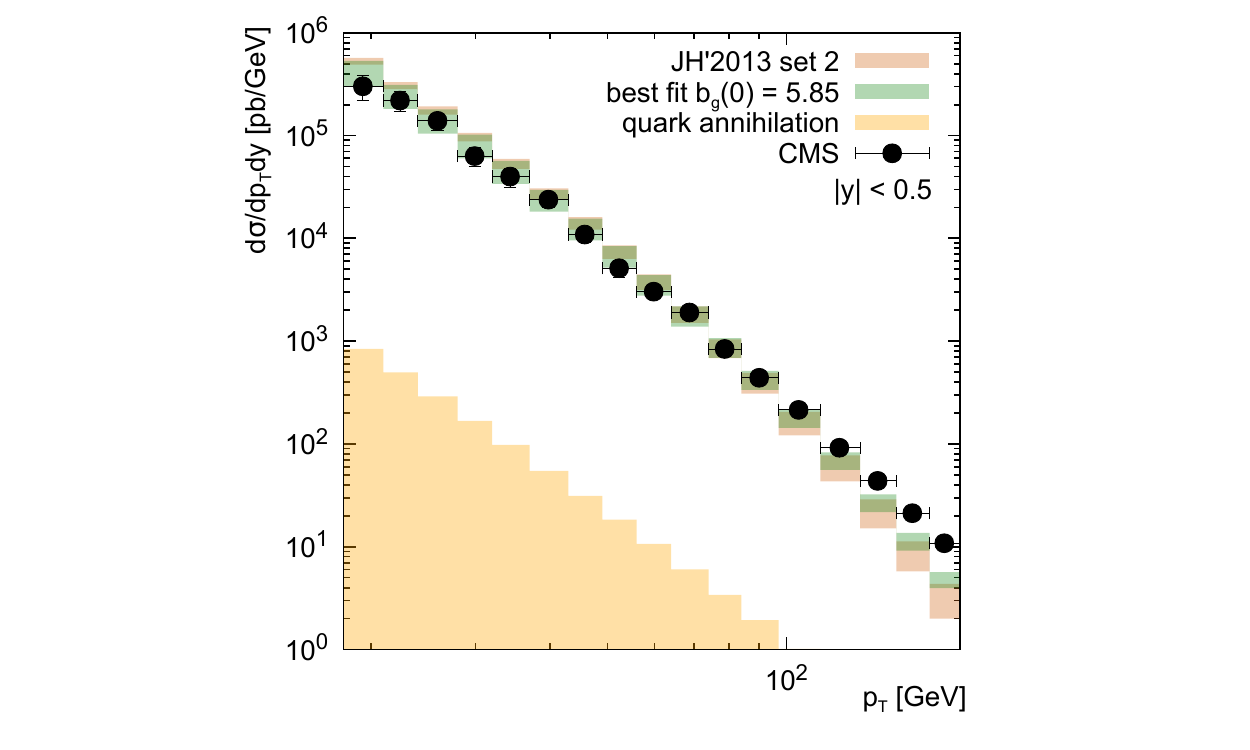}
\includegraphics[width=7.9cm]{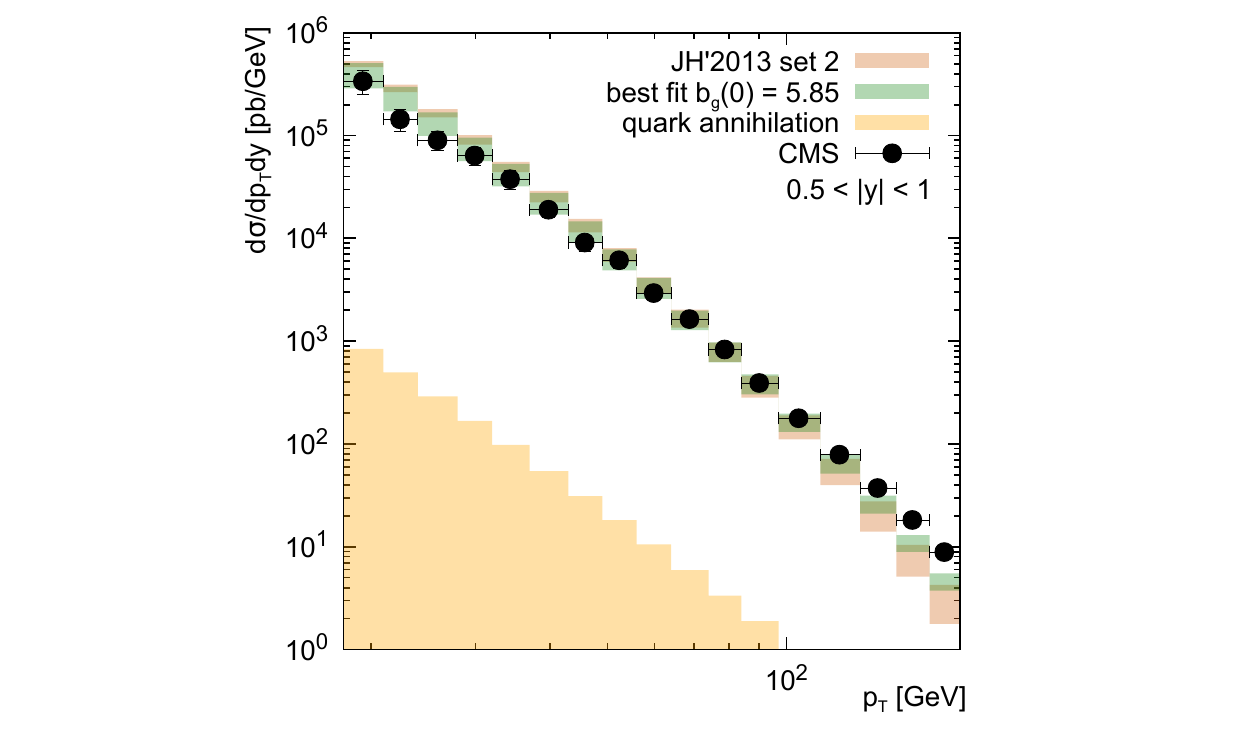}
\includegraphics[width=7.9cm]{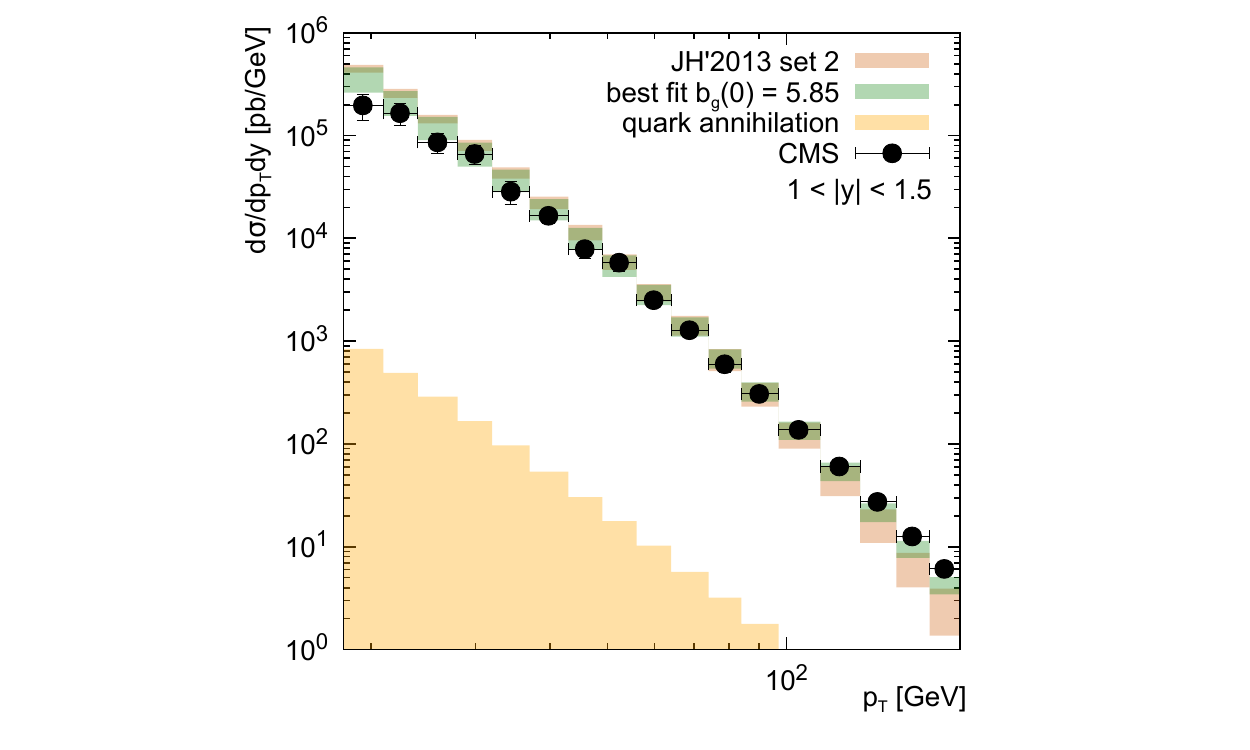}
\includegraphics[width=7.9cm]{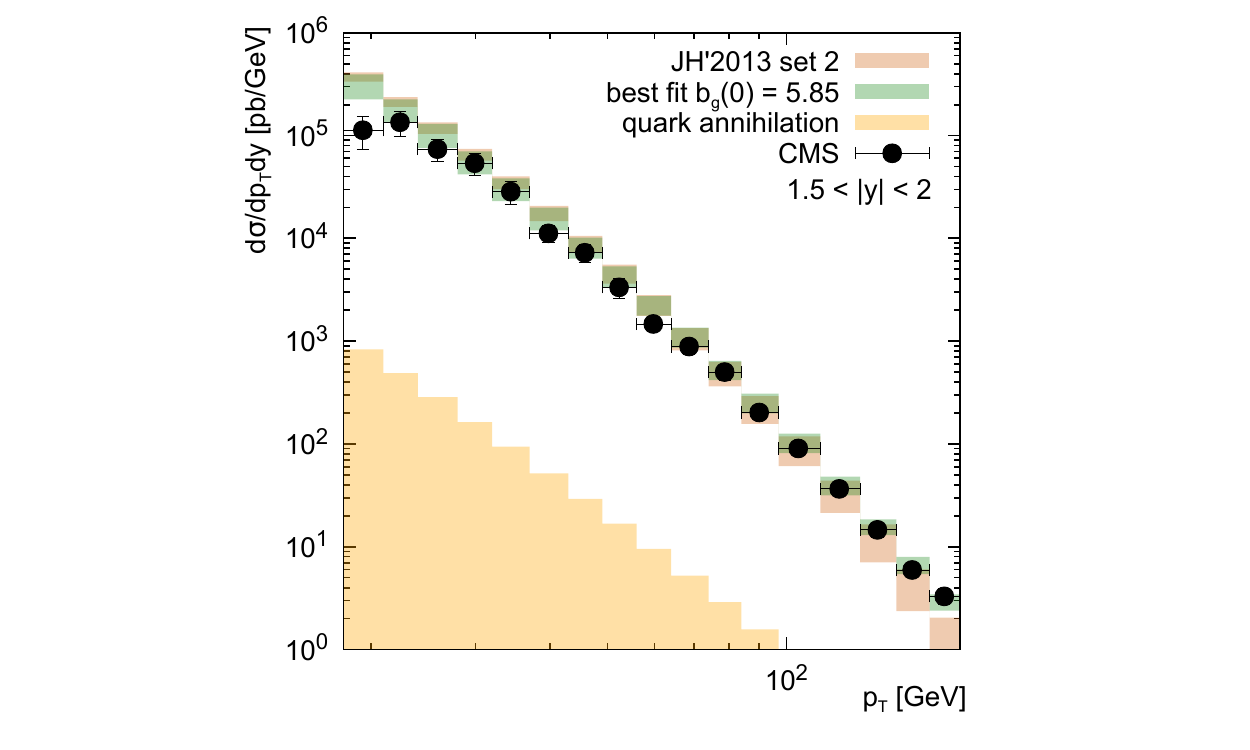}
\includegraphics[width=7.9cm]{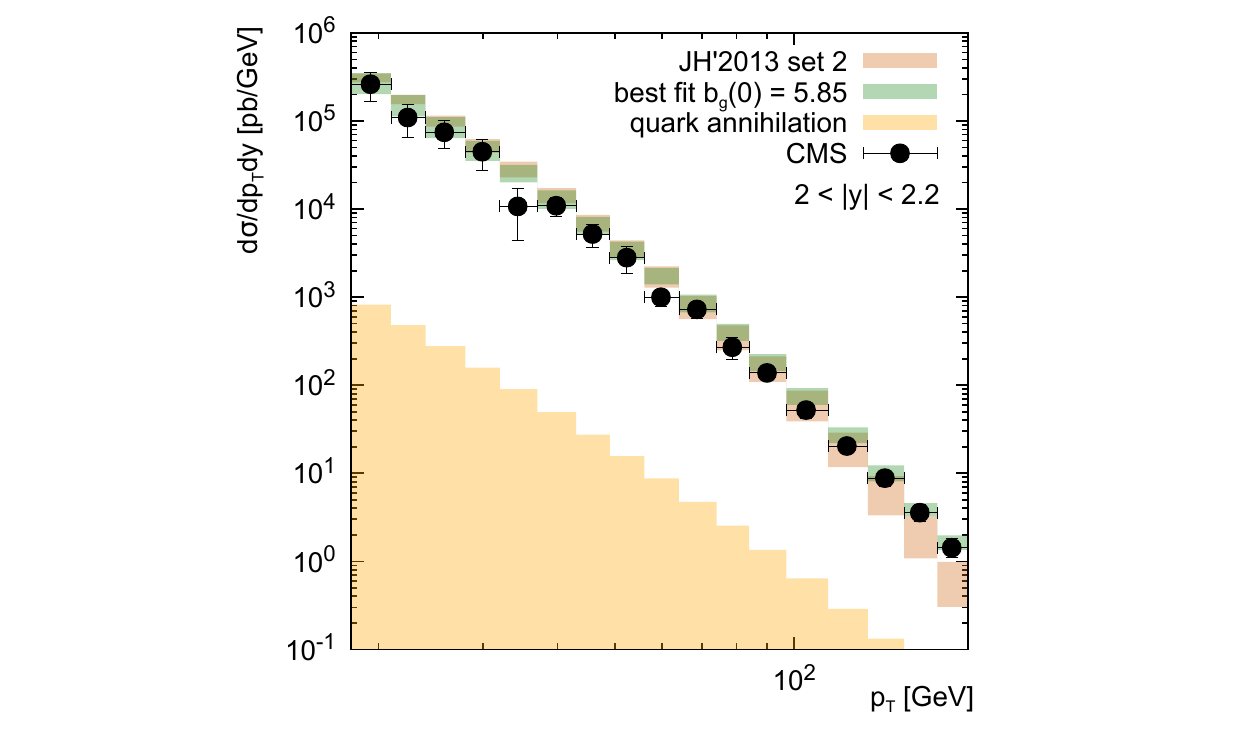}
\caption{Transverse momentum distributions of inclusive $b$-jets 
produced in $pp$ collisions at $\sqrt s = 7$~TeV at different rapidities
calculated using the CCFM-evolved TMD gluon density~(\ref{eg:OurGluon}) with fitted value $b_g(0) = 5.85$. 
Predictions obtained with the JH'2013 set 2 gluon are shown 
for comparison.
Shaded bands represent the estimation of theoretical uncertainties of our calculations.
Kinematical cuts are described in the text.
Experimental data are from CMS\cite{bjet-CMS-7}.}
\label{fig:fitbbCMS}
\end{center}
\end{figure}

The TMD gluon densities in a proton calculated with fitted value of $b_g(0)$
and $c_g$ 
are shown in Figs.~\ref{fig:Gluonx} and~\ref{fig:Gluonkt} 
as functions of proton's longitudinal momentum fraction $x$ and gluon transverse momentum ${\mathbf k}_T^2$
for different values of hard scale $\mu^2$. The shaded bands 
represent the uncertainties of our fitting procedure.
As one can see, these uncertainties become important at $x \geq 10^{-1}$.
For comparison, we also show the CCFM-evolved TMD gluon distributions from\cite{JH2013}, 
namely, JH'2013 set 2, which is often used in the
phenomenological applications.
In contrast with our approach, the $x$ dependence of JH'2013 set 2 input
has a general form given by~(\ref{eq:Jung-Hautmann}) with 
parameters derived from the 
high-precision HERA data on the proton structure functions $F_2(x, Q^2)$ and 
$F_2^c(x, Q^2)$ at $x < 5 \cdot 10^{-3}$ and $Q^2 > 3.5$~GeV$^2$.
We find that both these gluon densities have a remarkably different 
$x$ and ${\mathbf k}_T^2$ behaviour, especially in the region of 
small ${\mathbf k}_T^2$, see Fig.~\ref{fig:Gluonkt}.
Some phenomenological consequences of the latter we demostrate here.

\begin{figure}
\begin{center}
\includegraphics[width=7.9cm]{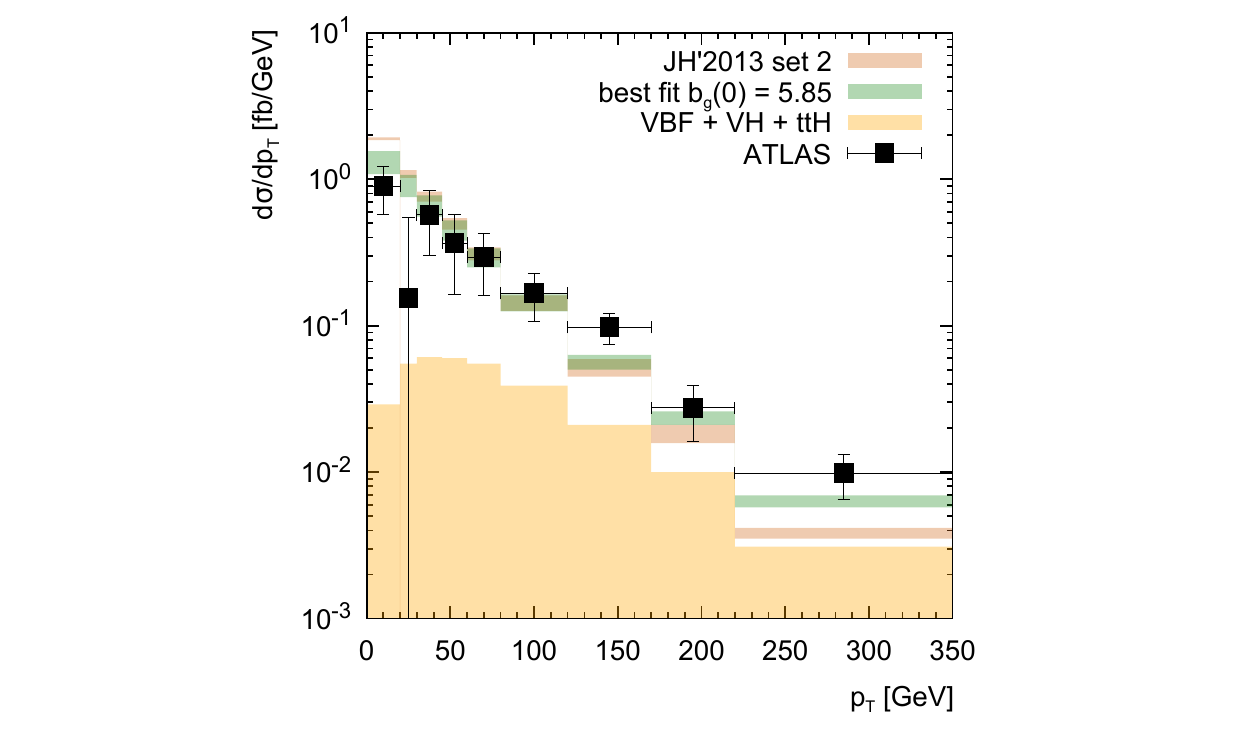}
\includegraphics[width=7.9cm]{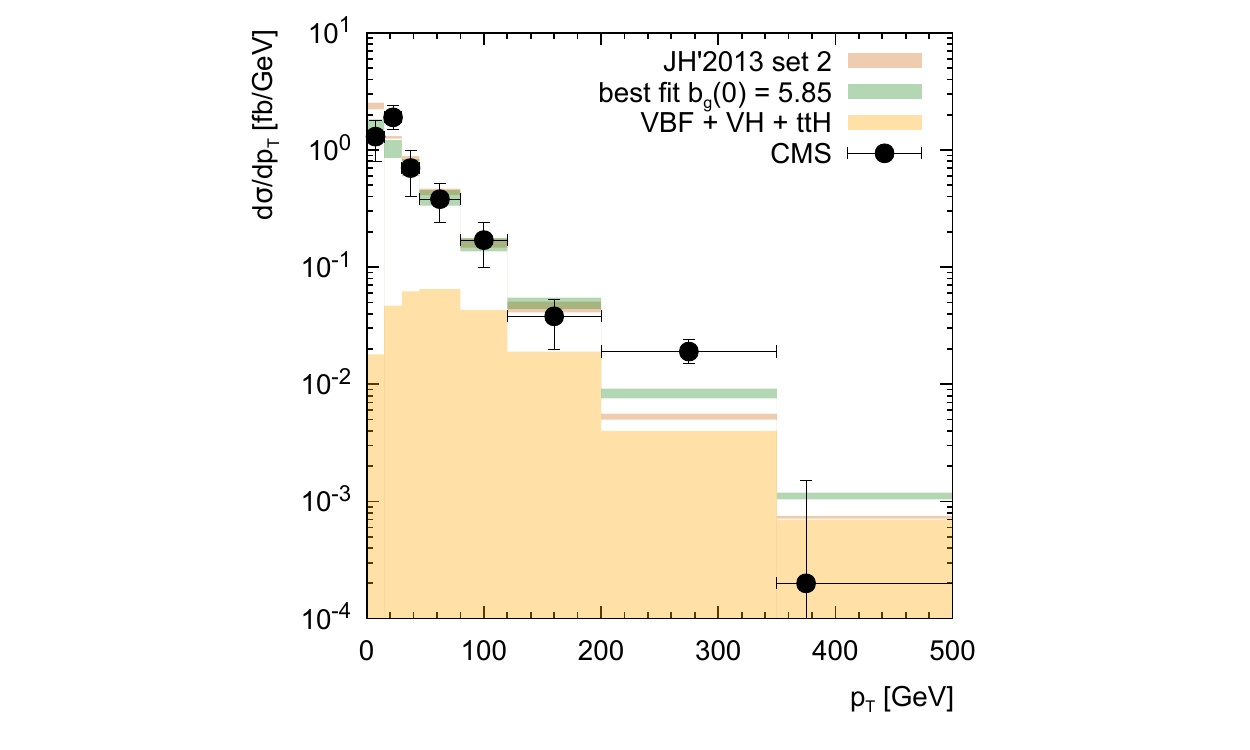}
\includegraphics[width=7.9cm]{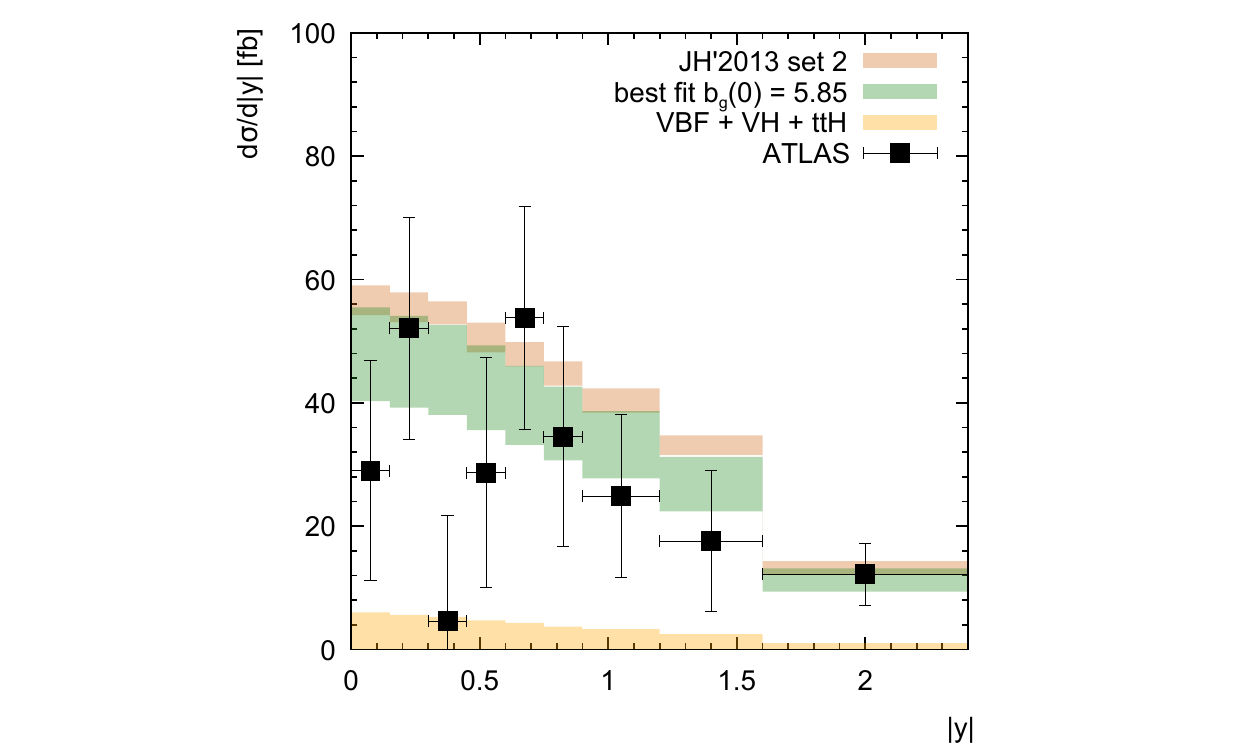}
\includegraphics[width=7.9cm]{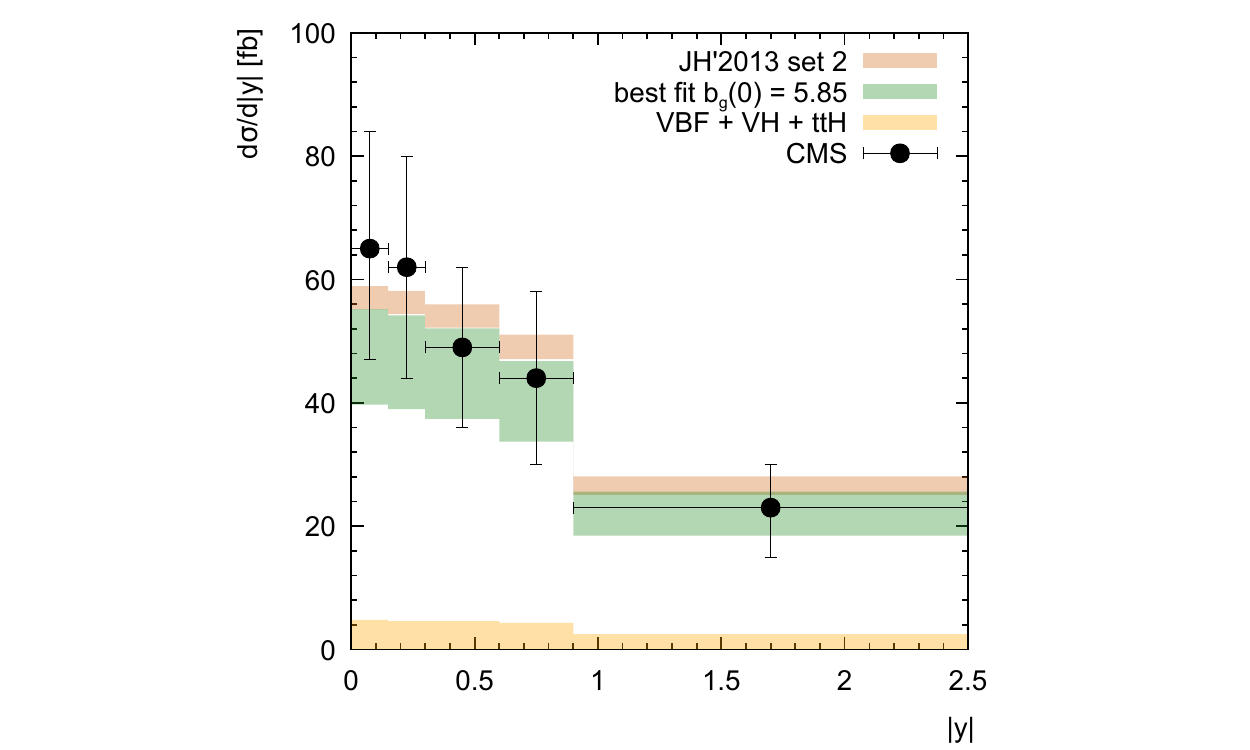}
\includegraphics[width=7.9cm]{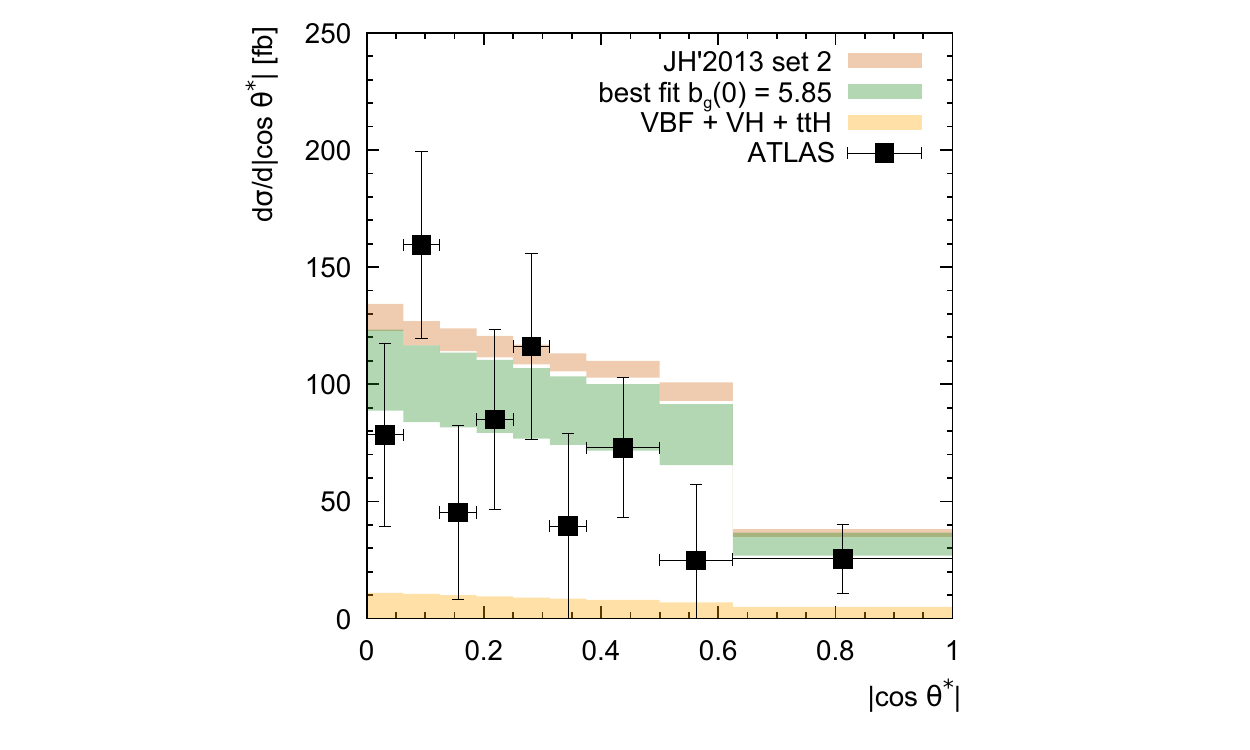}
\includegraphics[width=7.9cm]{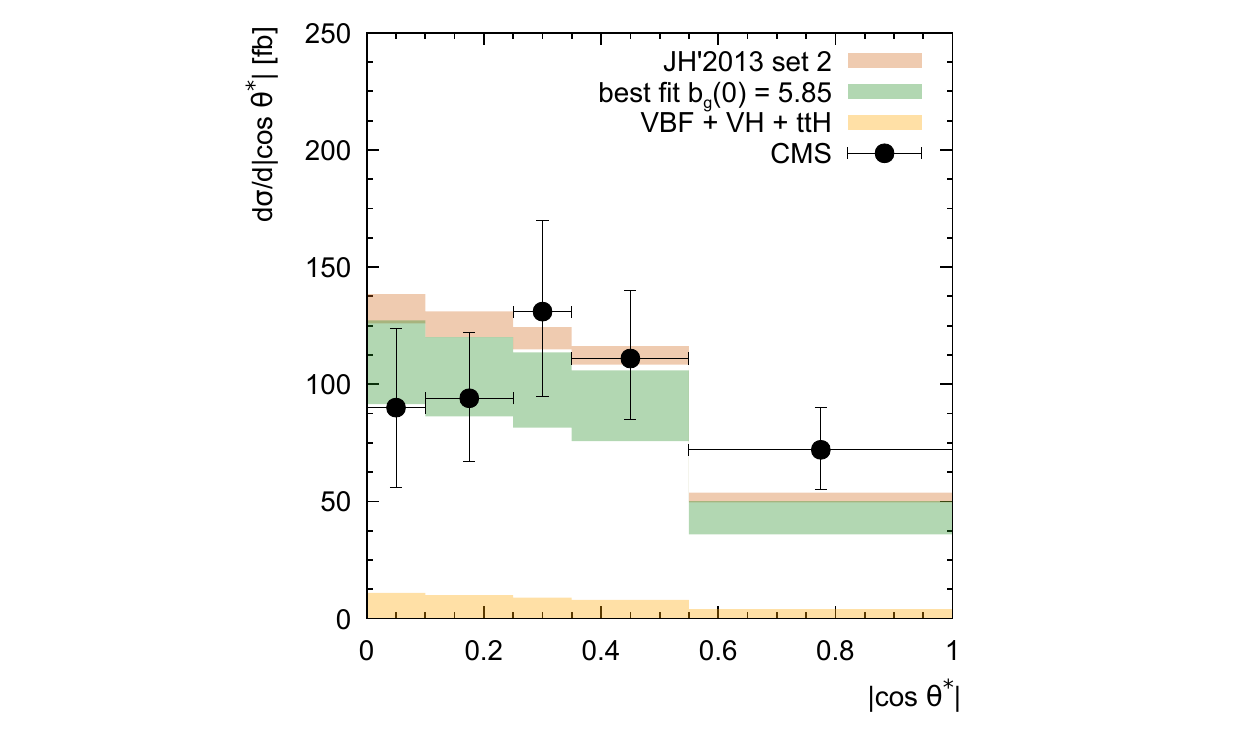}
\caption{Differential cross sections of inclusive Higgs boson production 
at $\sqrt s = 13$~TeV (in the diphoton decay mode) calculated as functions 
of diphoton transverse momentum $p_T$, rapidity $y$ and photon helicity angle $|\cos \theta^*|$
(in the Collins-Soper frame). Notation of all histograms is the same as in Fig.~\ref{fig:fitbbCMS}.
Kinematical cuts are described in the text.
Experimental data are from ATLAS\cite{HVV-ATLAS-13} and CMS\cite{HVV-CMS-13}.}
\label{fig:fitHiggsVV}
\end{center}
\end{figure}

\begin{figure}
\begin{center}
\includegraphics[width=7.9cm]{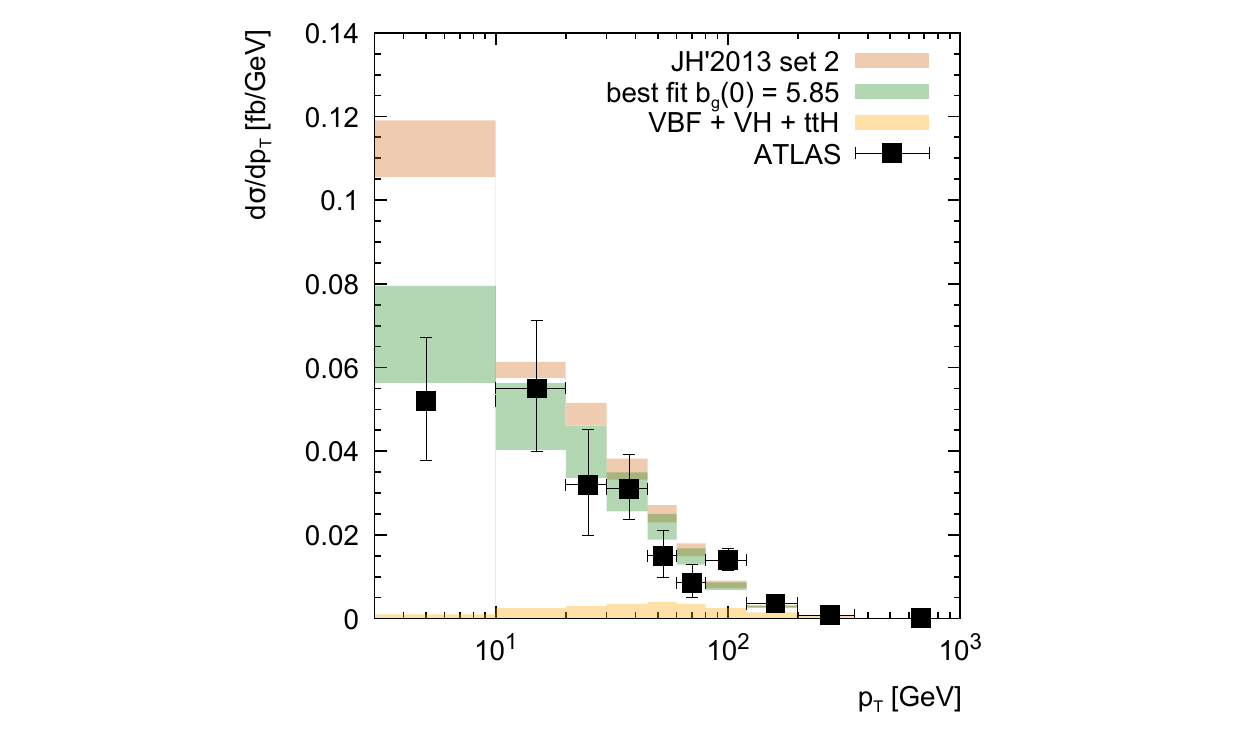}
\includegraphics[width=7.9cm]{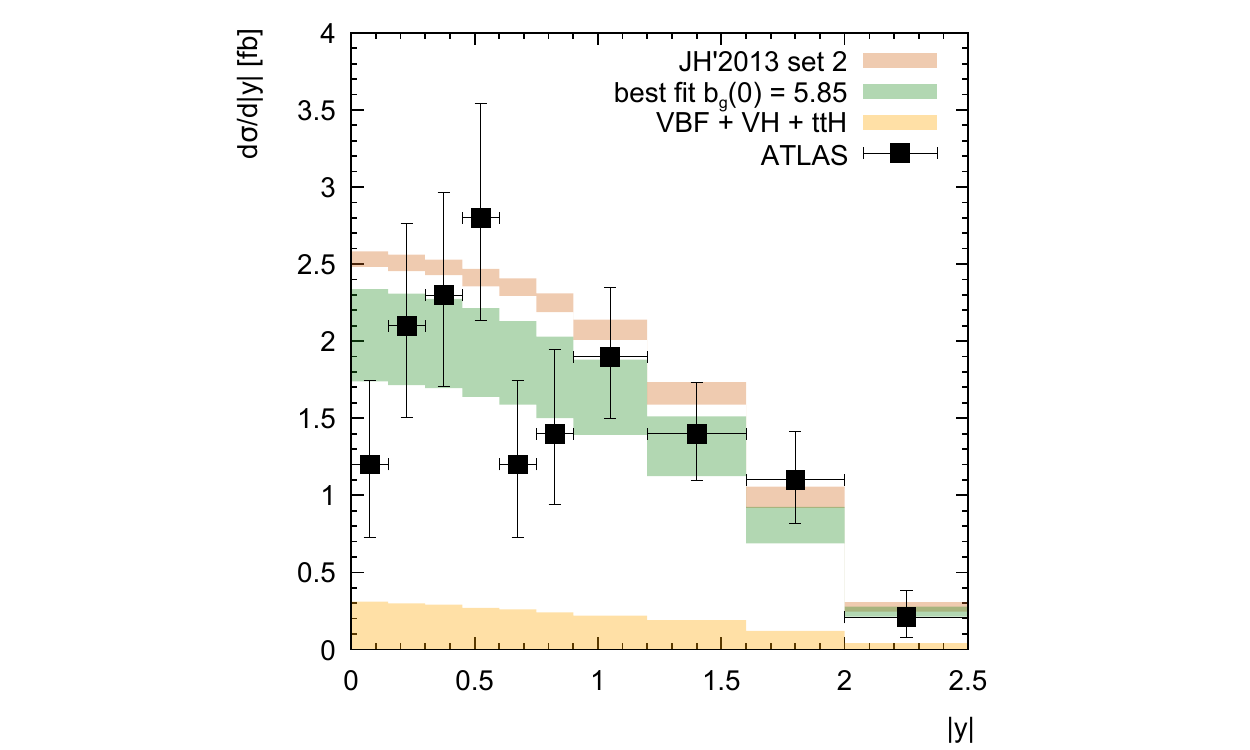}
\includegraphics[width=7.9cm]{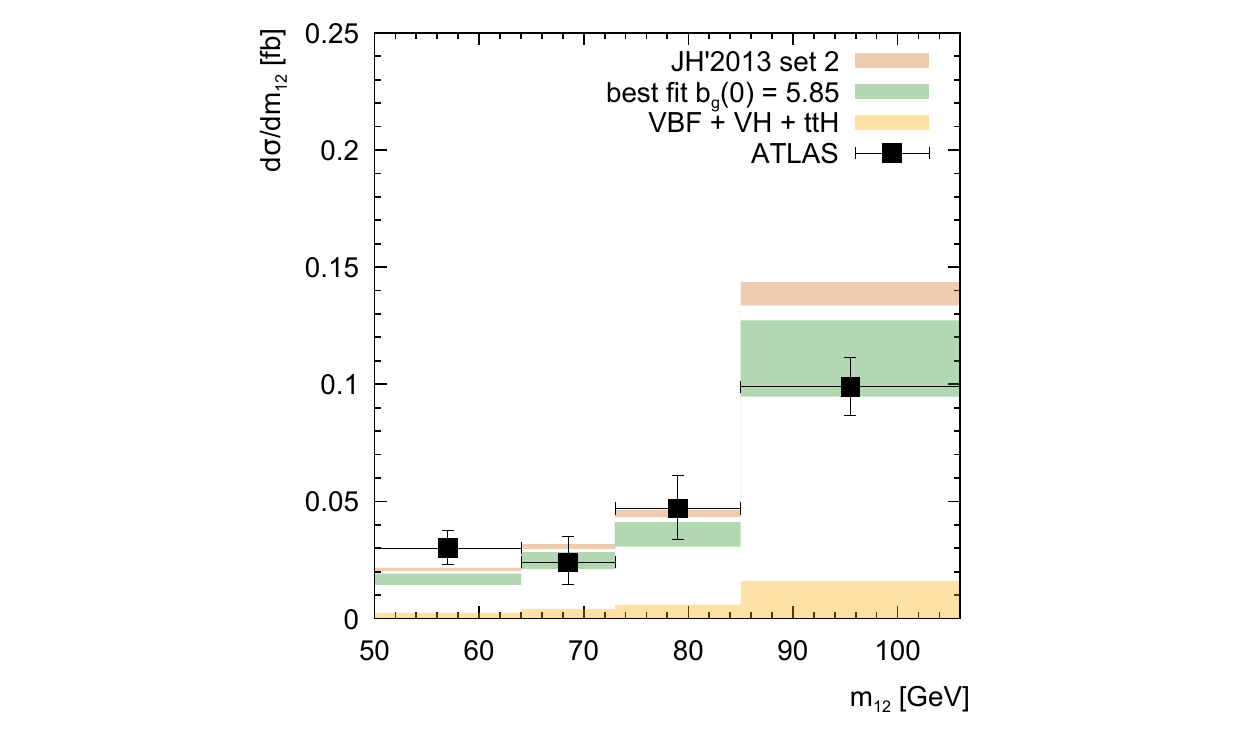}
\includegraphics[width=7.9cm]{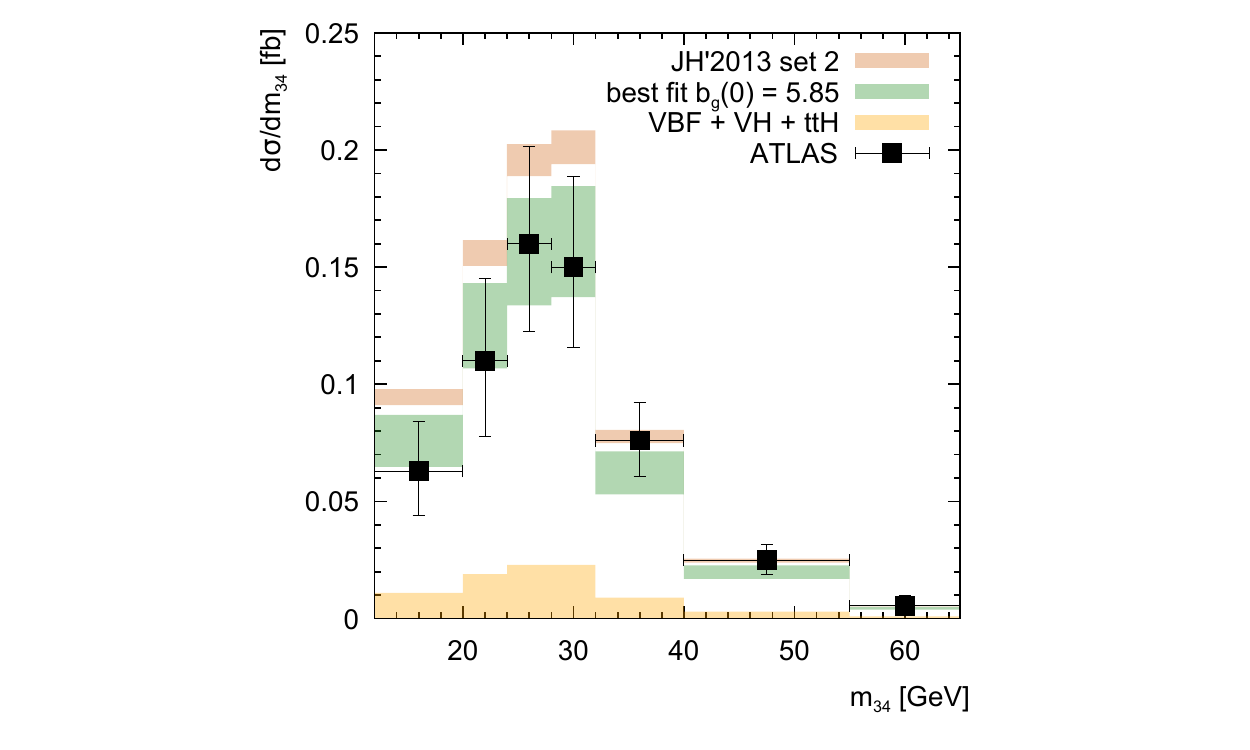}
\includegraphics[width=7.9cm]{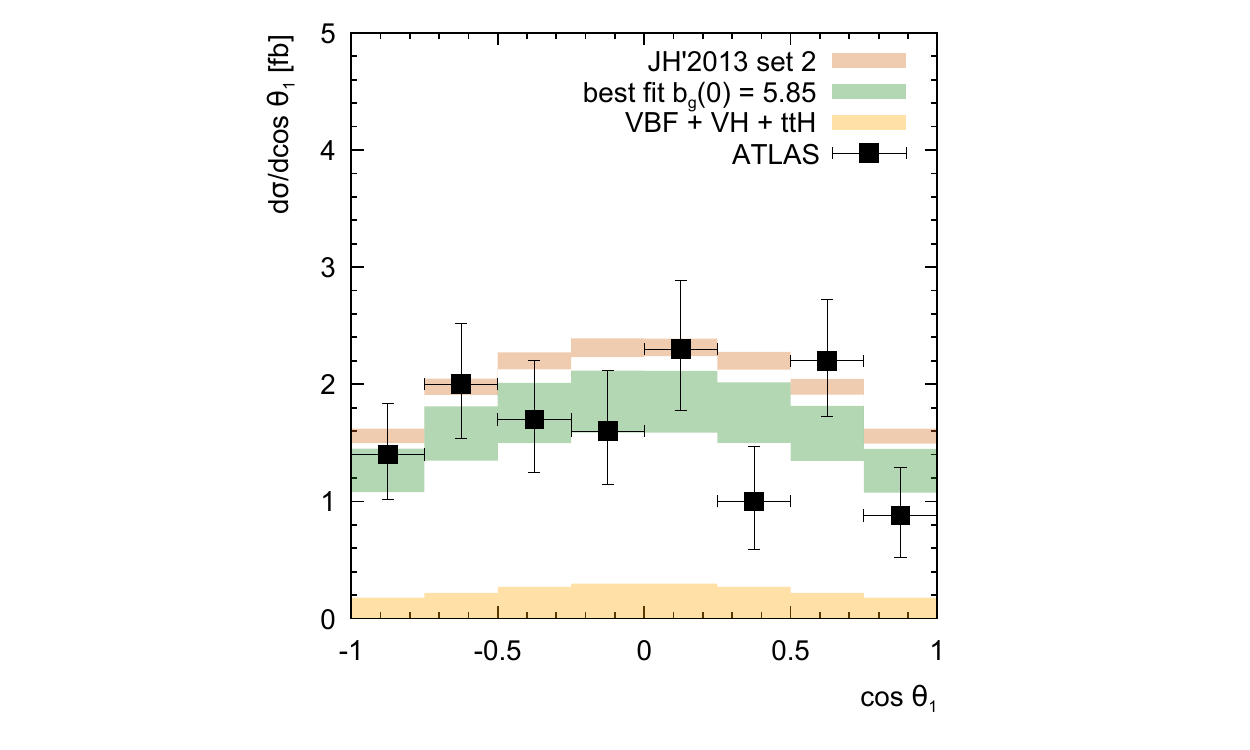}
\includegraphics[width=7.9cm]{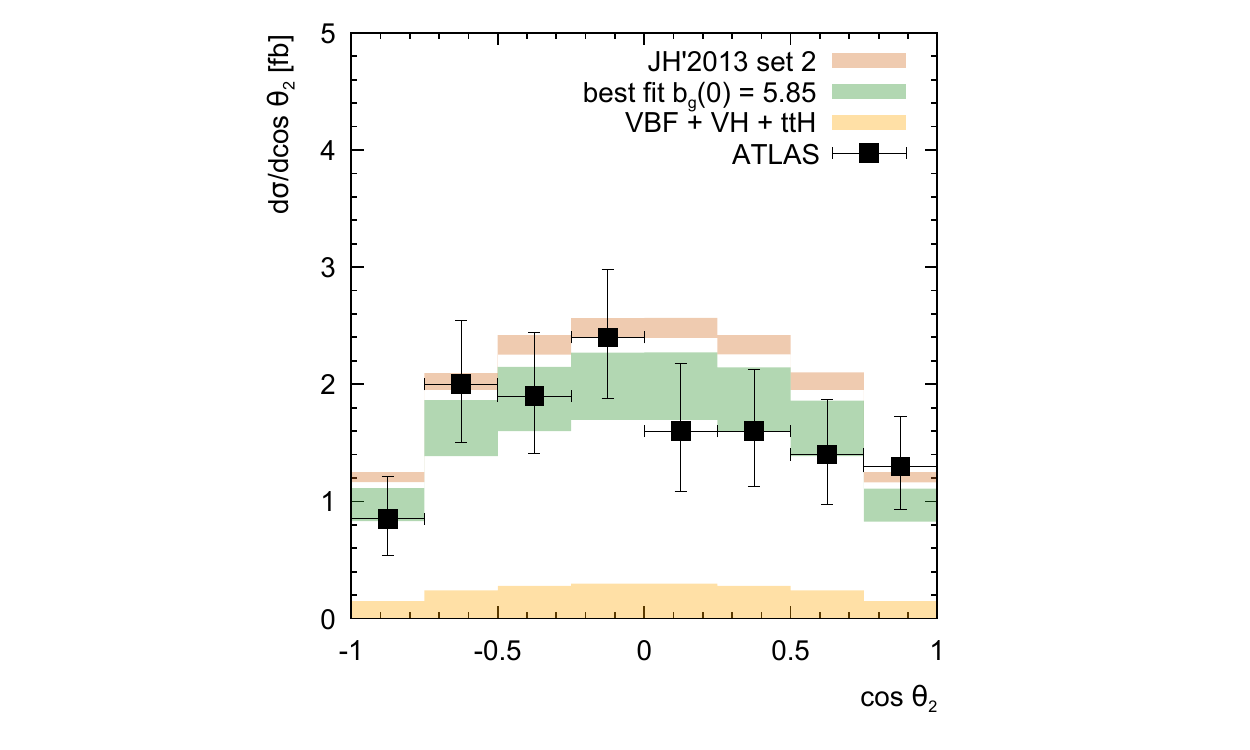}
\includegraphics[width=7.9cm]{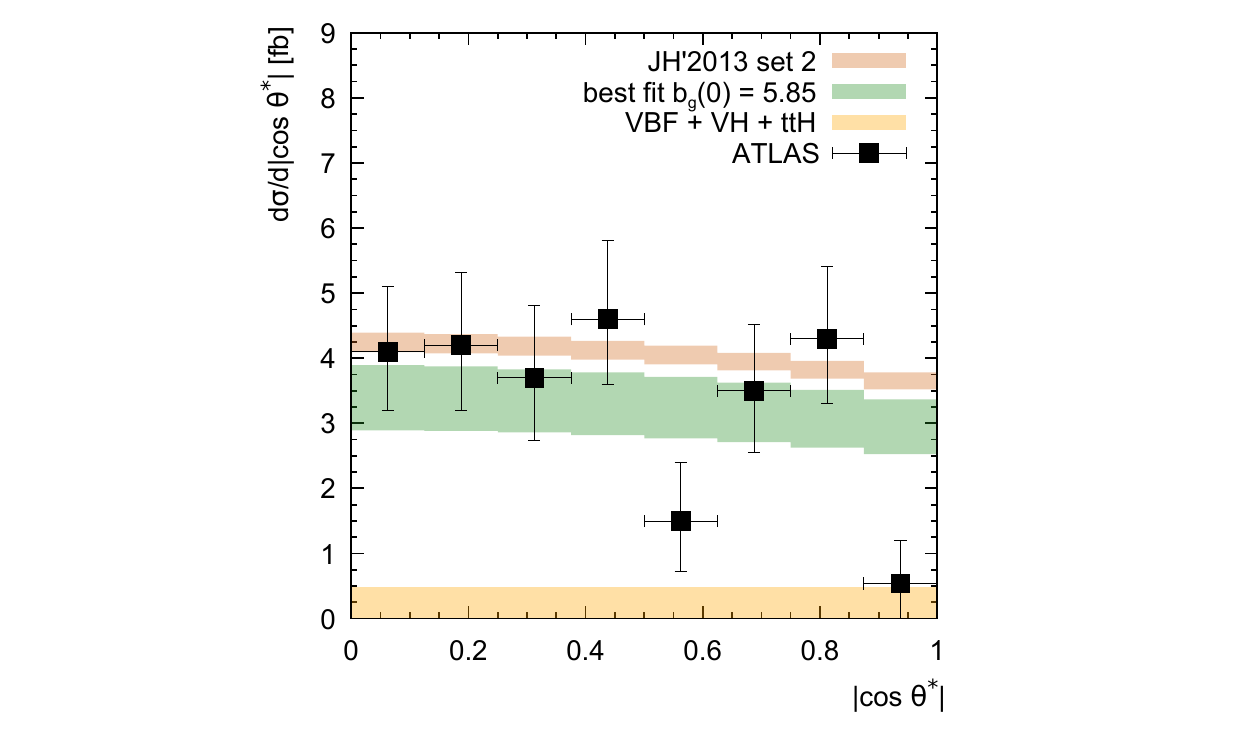}
\caption{Differential cross sections of inclusive Higgs boson production 
at $\sqrt s = 13$~TeV (in the $H \to ZZ^* \to 4l$ decay mode) calculated as functions 
of Higgs transverse momentum $p_T$, rapidity $y$, leading and subleading 
lepton pair invariant masses $m_{12}$ and $m_{34}$, leading lepton pair 
scattering angle $|\cos \theta^*|$ (in the Collins-Soper frame), 
first and second anti-lepton production
angles $\cos \theta_1$ and $\cos \theta_2$.
Notation of all histograms is the same as in Fig.~\ref{fig:fitbbCMS}.
Kinematical cuts are described in the text.
Experimental data are from ATLAS\cite{HZZ-ATLAS-13}.}
\label{fig:fitHiggsZZ}
\end{center}
\end{figure}

\begin{figure}
\begin{center}
\includegraphics[width=15cm]{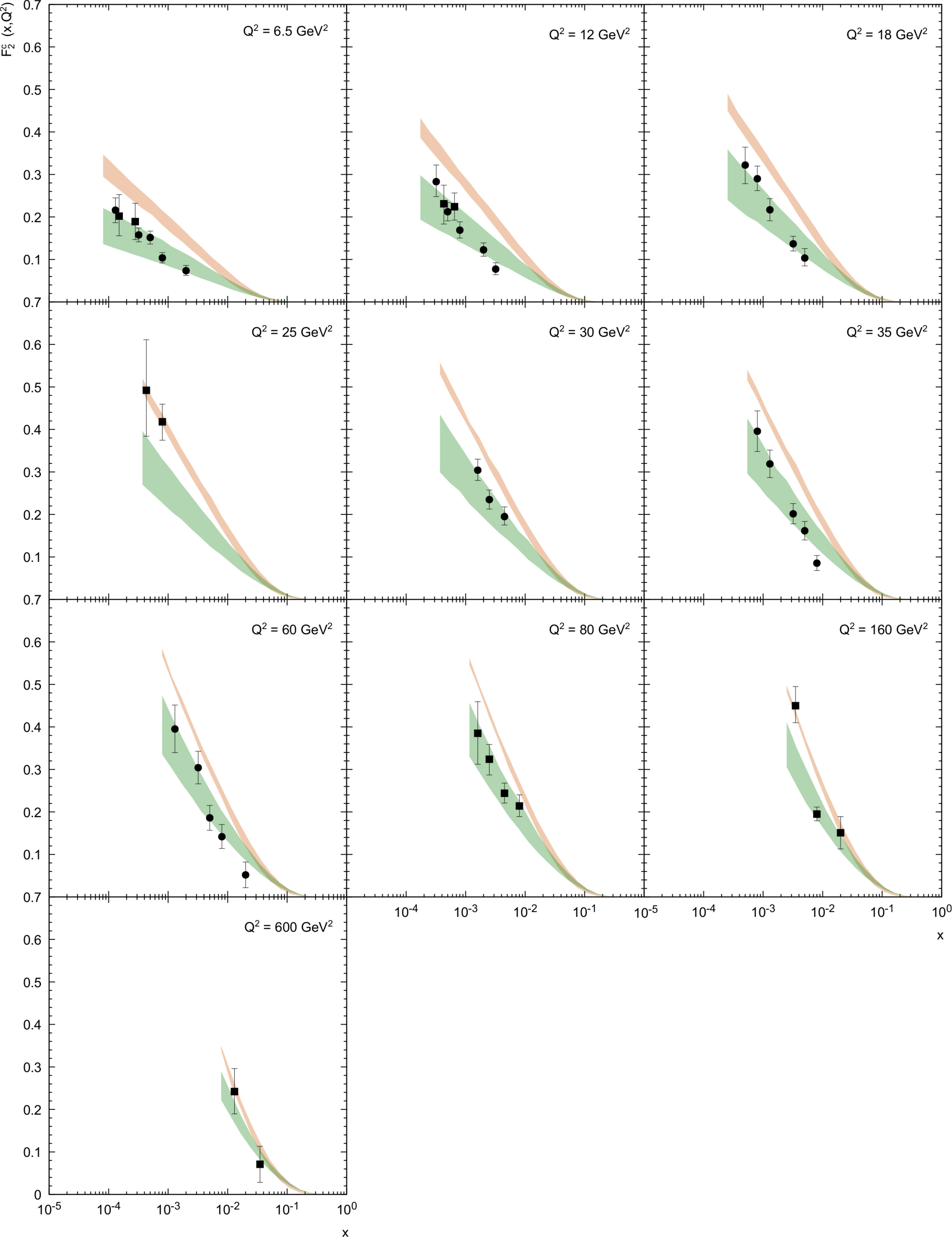}
\caption{Structure functions $F_2^c(x,Q^2)$ measured at different scales
calculated using the CCFM-evolved TMD gluon density~(\ref{eg:OurGluon}) with fitted value $b_g(0) = 5.85$. 
Predictions obtained with the JH'2013 set 2 gluon are shown 
for comparison.
Shaded bands represent the estimation of theoretical uncertainties of our calculations.
Experimental data are from ZEUS and H1\cite{F2cb-ZEUS, F2c-H1, F2cb-H1}.}
\label{fig:fitFc}
\end{center}
\end{figure}

\begin{figure}
\begin{center}
\includegraphics[width=15cm]{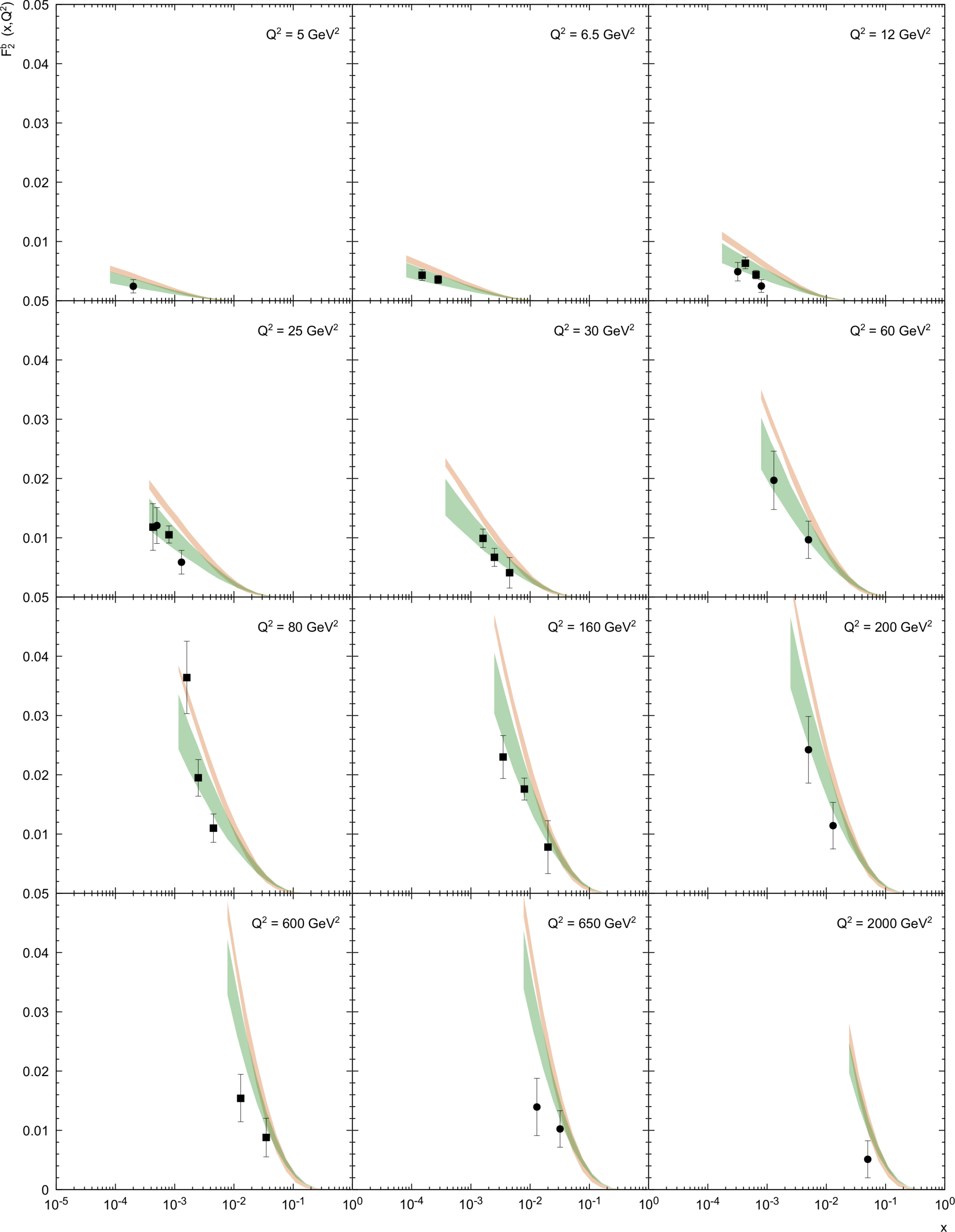}
\caption{Structure functions $F_2^b(x,Q^2)$ measured at different scales
calculated using the CCFM-evolved TMD gluon density~(\ref{eg:OurGluon}) with fitted value $b_g(0) = 5.85$. 
Predictions obtained with the JH'2013 set 2 gluon are shown 
for comparison.
Shaded bands represent the estimation of theoretical uncertainties of our calculations.
Experimental data are from ZEUS and H1\cite{F2cb-ZEUS, F2cb-H1}.}
\label{fig:fitFb}
\end{center}
\end{figure}

\begin{figure}
\begin{center}
\includegraphics[width=15cm]{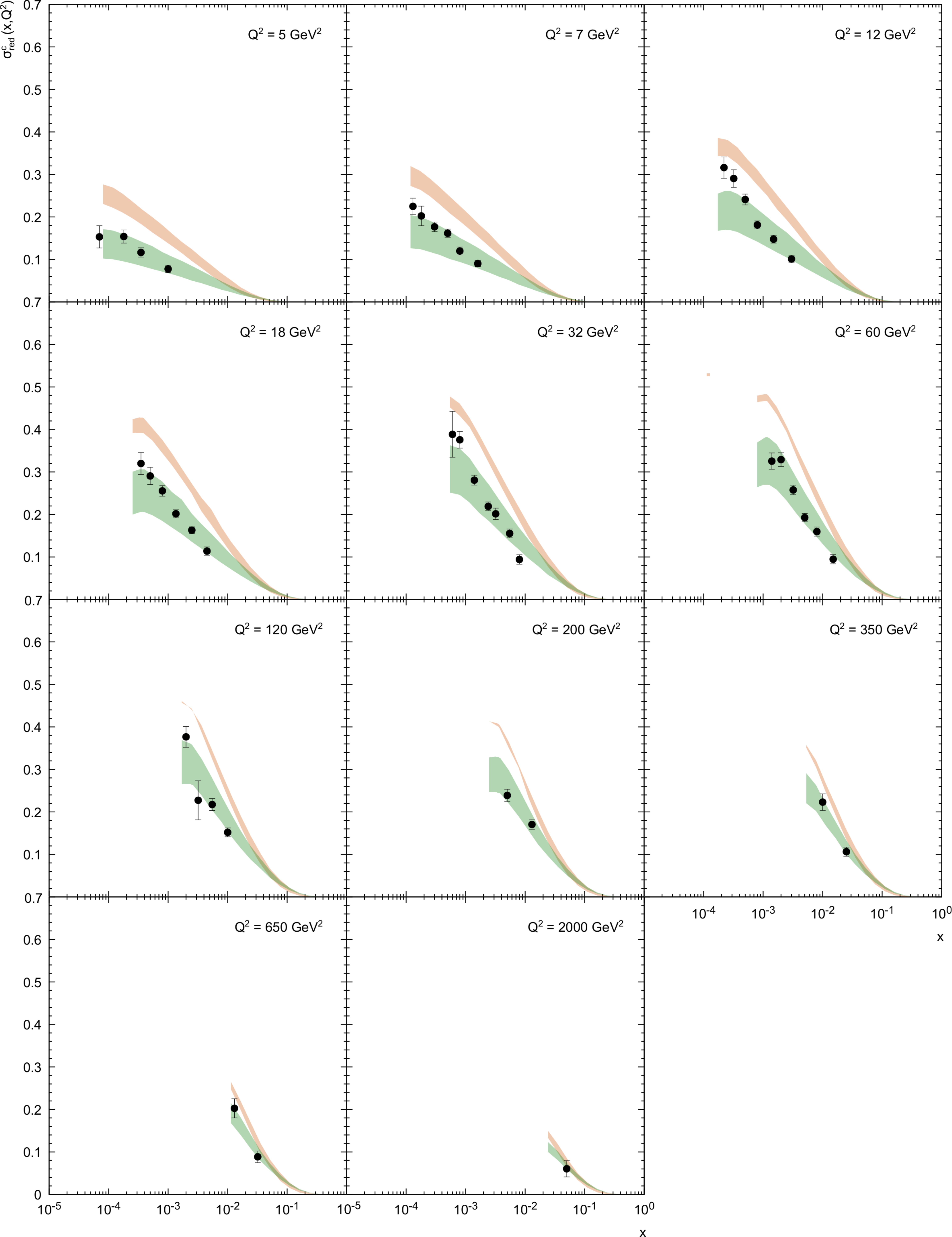}
\caption{Reduced cross sections $\sigma_\text{red}^{c}(x,Q^2)$ measured at different scales
calculated using the CCFM-evolved TMD gluon density~(\ref{eg:OurGluon}) with fitted value $b_g(0) = 5.85$. 
Predictions obtained with the JH'2013 set 2 gluon are shown 
for comparison.
Shaded bands represent the estimation of theoretical uncertainties of our calculations.
Experimental data are from ZEUS and H1~\cite{sigma_red-ZEUS+H1}.}
\label{fig:fitsigmaredc}
\end{center}
\end{figure}

\begin{figure}
\begin{center}
\includegraphics[width=15cm]{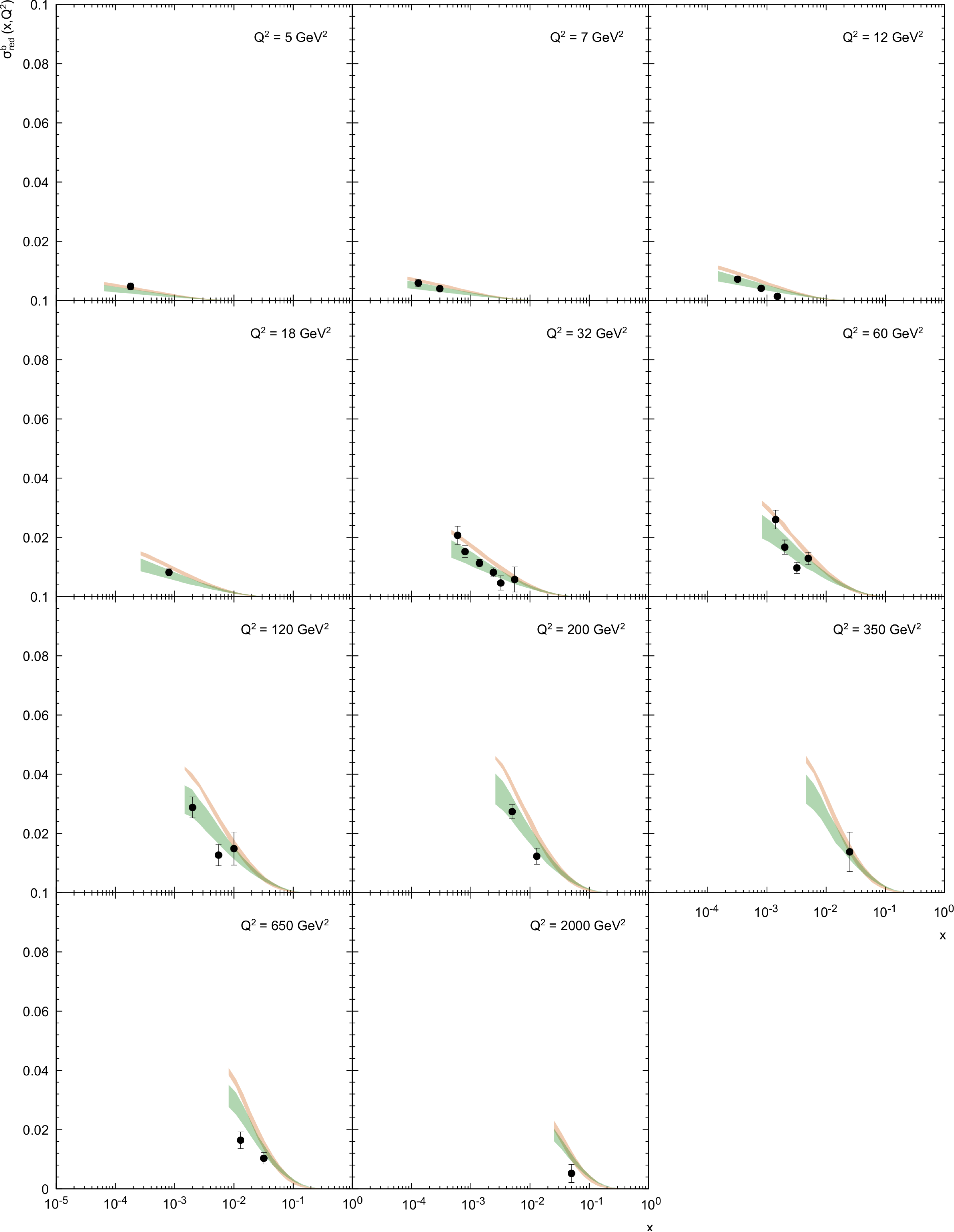}
\caption{Structure functions $\sigma_\text{red}^{b}(x,Q^2)$ measured at different scales
calculated using the CCFM-evolved TMD gluon density~(\ref{eg:OurGluon}) with fitted value $b_g(0) = 5.85$. 
Predictions obtained with the JH'2013 set 2 gluon are shown 
for comparison.
Shaded bands represent the estimation of theoretical uncertainties of our calculations.
Experimental data are from ZEUS and H1~\cite{sigma_red-ZEUS+H1}.}
\label{fig:fitsigmaredb}
\end{center}
\end{figure}

So, the experimental data involved in our fit
are compared with the obtained predictions in Figs.~\ref{fig:fitbbCMS} --- \ref{fig:fitsigmaredb}. 
The shaded bands represent the
theoretical uncertainties of our calculations.
For comparison, we also used here the JH'2013 set 2 gluon distribution.
One can see that our fit leads to a good agreement with the
experimental data practically for all considered observables.
The HERA data on structure functions $F_2^c(x, Q^2)$, $F_2^c(x, Q^2)$ and 
reduced cross sections $\sigma_{\rm red}^c(x, Q^2)$, $\sigma_{\rm red}^b(x, Q^2)$
are reasonably well described by both considered TMD gluons
within the uncertainties.
However, we find that the JH'2013 set 2 gluon density provides a bit worse
description of $b$-jet and/or Higgs boson production at the LHC,
especially at low 
transverse momenta (see Figs.~\ref{fig:fitbbCMS} --- \ref{fig:fitHiggsZZ}).
The better agreement of these data 
achieved with the proposed TMD gluon density 
is an immediate consequense of using the physically motivated expression~(\ref{eg:OurGluon}) for 
input distribution. 
In fact, at low 
transverse momenta 
the relative small gluon ${\mathbf k}_T^2$ are probed, where the 
difference between the considered gluon distributions becomes essential, as it is shown in Fig.~\ref{fig:Gluonkt}.
Moreover, significant overestimation of the measured $b$-jet and especially Higgs boson 
$p_T$-spectra at low $p_T$ obtained with JH'2013 set~2 gluon leads to a notable 
difference in absolute normalization of Higgs rapidity, decay 
photon scattering angle $\cos \theta^*$, invariant masses $m_{12}$, $m_{34}$
and other observables shown in Figs.~\ref{fig:fitHiggsVV} and \ref{fig:fitHiggsZZ}.
So, our calculations clearly demonstrate
that experimental data for considered processes are
strongly sensitive to the TMD gluon distribution in the proton
and can be used to constrain the latter.
Of course, future more precise measurements could be very useful and important 
to reduce uncertainties in determination of the phenomenological 
parameters from the data.

We performed here a step forward to the global analysis of 
collider data in the $k_T$-factorization approach.
The obtained TMD gluon density is already available for the community. 
It is implemented into the Monte-Carlo event 
generator \textsc{pegasus}. Moreover, it is included in 
\textsc{tmdlib} package, which is used by the 
\textsc{cascade} and \textsc{katie} Monte-Carlo generators. 

\section{Conclusion} \indent

We have proposed a new analytical expression for the TMD
gluon density in the proton valid in a soft kinematical region. Using the modified quark-quark gluon string model,
where gluonic state and non-zero transverse momentum of partons inside the proton
are taken into account, we have obtained some 
corresponding phenomenological parameters
from the best description of LHC data on charged hadron (pion and kaon) spectra produced in $pp$ collisions 
at low transverse momenta $p_T \simeq 1$~GeV.
We have shown that the new suggested TMD gluon distribution 
incorporates saturation effects for the dipole cross section 
at a scale lower than the prediction of GBW model.
Then, treating the obtained TMD gluon distribution as the initial 
condition for the subsequent non-collinear QCD evolution, we have extended it 
to the whole kinematical region using the CCFM equation.
Several parameters important at moderate and large $x$ have been fitted from the LHC data on 
inclusive $b$-jet and Higgs boson production as well as latest HERA data on
proton structure functions $F_2^c(x, Q^2)$ and $F_2^b(x, Q^2)$ and reduced 
cross sections $\sigma_{\rm red}^c(x, Q^2)$ and $\sigma_{\rm red}^b(x, Q^2)$. 
Our fit leads to simultaneous description of all these processes with good $\chi^2 /d.o.f. = 2.2$. 
The obtained TMD gluon distribution in a proton is available for 
public usage and implemented in the popular \textsc{tmdlib} package and 
Monte-Carlo event generator \textsc{pegasus}.

\section*{Acknowledgements} \indent

We thank S.P.~Baranov, A.A.~Prokhorov, H.~Jung, S.~Turchikhin and S.~Taheri Monfared 
for their important comments and remarks. 
Studies described in Section~2 were supported by the 
Russian Science Foundation under grant~22-22-00387. 
Updated release of the Monte-Carlo generator \textsc{pegasus} (version 1.07.01)
was supported by the Russian Science Foundation, grant~22-22-00119.

\section*{Appendix A} \indent

In the conventional QGSM framework (neglecting the transverse momentum dependence)
the quark and diquark distribution functions in a proton $f_a(x)$, where 
$a = u$, $d$, $ud$ or $uu$, were calculated\cite{SoftQuarkGluonStringModel-1, SoftQuarkGluonStringModel-2}.
They read:
$$
  f_u(x) = C_u^p x^{-1/2}(1-x)^{3/2}, \quad f_d(x) = C_d^p x^{-1/2}(1-x)^{5/2}, \eqno(\text{A1})
$$
$$
  f_{ud}(x) = C_{ud}^p (1 -x)^{-1/2}x^{3/2}, \quad f_{uu}(x) = C_{ud}^p (1 -x)^{-1/2}x^{5/2}. \eqno(\text{A2})
$$
\noindent
Here overall normalization factors are given by
$$
  C_u^p = C_{ud}^p = \frac{\Gamma(2-1/2+3/2)}{\Gamma(1-1/2)\Gamma(1+3/2)} = 1/1.1781,\eqno(\text{A3})
$$
$$
  C_d^p = C_{uu}^p = \frac{\Gamma(2-1/2+5/2)}{\Gamma(1-1/2)\Gamma(1+5/2)}=1/1.01859.\eqno(\text{A4})
$$
\noindent
In the modified QGSM, where gluonic state in the proton and 
partonic transverse momentum are taken into account, the TMD quark and diquark densities 
can be written as
$$
  f_a(x, {\mathbf k}_T^2) = c_a f_a(x) g_a({\mathbf k}_T^2), \quad g_{a}({\mathbf k}_{T}^2)= \frac{B_{a}^2}{2\pi}e^{ -B_{a} |{\mathbf k}_{T}|},\eqno(\text{A5})
$$
\noindent
where $B_q = B_{qq} = 5$~GeV$^{-1}$. The normalization factors $c_a$ (which are about of $1/2$) 
can be included to restore QCD quark-gluon sum rules.

Here we list analytical expressions for quark, diquark and gluon 
fragmentation functions used in our calculations. These expressions were
obtained in the QGSM at the LO\cite{FFs}.
So, for gluons we have
$$
  G_{g\to h}(z,|{\mathbf p}_T|) = 2 G_{g\to\pi}(z) I^g_{\pi}(|{\mathbf p}_T|) + 2 G_{g\to K}(z) I^g_{K}(|{\mathbf p}_T|),\eqno(\text{A6})
$$
\noindent
where the coefficients $2$ come from the following relations:
$$
  G_{g\to\pi^+}(z) = G_{g\to\pi^-}(z), \quad G_{g\to K^+}(z) = G_{g\to K^-}(z).\eqno(\text{A7})
$$
\noindent
The parametrizations of $G_{g\to \pi} (z)$ and $G_{g\to K} (z)$ are the following:
$$
  G_{g\to \pi} (z) = 6.57 z^{0.54}  (1-z)^{3.01},\eqno(\text{A8})
$$
$$
  G_{g\to K} (z)  = 0.37 z^{0.79}  (1-z)^{3.07},\eqno(\text{A9})
$$
\noindent
and functions $I^g_{\pi}(|{\mathbf p}_T|)$ and $I^g_{K}(|{\mathbf p}_T|)$ read:
$$
  I^g_{\pi}(|{\mathbf p}_T|) = I^g_{K}(|{\mathbf p}_T|) = I^g_{h}(|{\mathbf p}_T|) = \frac{(B^g_{f_{h}})^2}{ 2\pi} e^{-B^g_{f_{h}} |{\mathbf p}_T|},\eqno(\text{A10})
$$
\noindent
with $B^g_{f_{h}} = B^g_{f_{\pi}} = B^g_{f_{K}} = 4.5$~GeV$^{-1}$.
For quark fragmentation functions we have:
$$
  G_{u\rightarrow \pi^+}(z,|{\mathbf p}_T|) = \left[a_0(1-z) + a_0(1-z)^2\right] I^q_{\pi}(|{\mathbf p}_T|),\eqno(\text{A11})
$$
$$
  G_{d\rightarrow \pi^+}(z, |{\mathbf p}_T|) =(1-z) G_{u\rightarrow \pi^+}(z) I^q_{\pi}(|{\mathbf p}_T|),\eqno(\text{A12})
$$
$$
  G_{u\rightarrow K^+}(z, |{\mathbf p}_T|) = a_k(1-z)^{1/2}(1+a_{1K}z) I^q_{K}(|{\mathbf p}_T|),\eqno(\text{A13})
$$
$$
  G_{u\rightarrow K^-}(z, |{\mathbf p}_T|) = a_k(1-z)^{3/2} I^q_{K}(|{\mathbf p}_T|),\eqno(\text{A14})
$$
$$
  G_{d\rightarrow K^+}(z, |{\mathbf p}_T|) = G_{u\rightarrow K^-}(z) I^q_{K}(|{\mathbf p}_T|),\eqno(\text{A15})
$$
$$
  G_{d\rightarrow K^-}(z, |{\mathbf p}_T|) = G_{u\rightarrow K^+}(z) I^q_{K}(|{\mathbf p}_T|).\eqno(\text{A16})
$$
\noindent
Analytical forms for $I^q_{\pi}(|{\mathbf p}_T|)$ and $I^q_{K}(|{\mathbf p}_T|)$ are the same 
as for $I^g_{\pi}(|{\mathbf p}_T|)$, $I^g_{K}(|{\mathbf p}_T|)$, but the 
slopes are $B^q_{f_{\pi}} = B^q_{f_{K}} = 7$~GeV$^{-1}$. 
Other parameters are $a_0=0.65$, $a_k=0.075$ and $a_{1K}=2$.
Finally, for diquarks one can write:
$$
  G_{uu \rightarrow \pi^+}(z, |{\mathbf p}_T|) =a_0(1-z)^{2} I^{qq}_{\pi}(|{\mathbf p}_T|),\eqno(\text{A17})
$$
$$
  G_{ud \rightarrow \pi^+}(z, |{\mathbf p}_T|)= a_0(1+(1-z)^2)(1-z)^{2}  I^{qq}_{\pi}(|{\mathbf p}_T|),\eqno(\text{A18})
$$
$$
  G_{uu\rightarrow K^+}(z, |{\mathbf p}_T|) = a_k(1-z)^{5/2}(1+a_{2K}z) I^{qq}_{K}(|{\mathbf p}_T|),\eqno(\text{A19})
$$
$$
 G_{uu\rightarrow K^-}(z, |{\mathbf p}_T|)= a_k(1-z)^{7/2} I^{qq}_{K}(|{\mathbf p}_T|),\eqno(\text{A20})
$$
$$
G_{ud\rightarrow K^+}(z, |{\mathbf p}_T|) =\frac{a_k}{2}  (1-z)^{5/2} (1+a_{2K}z+ (1-z)^2) I^{qq}_{K}(|{\mathbf p}_T|),\eqno(\text{A21})
$$
$$
 G_{ud\rightarrow K^-}(z, |{\mathbf p}_T|)=\frac{a_k}{2}(1-z)^{7/2} (1 + (1-z)^2) I^{qq}_{K}(|{\mathbf p}_T|),\eqno(\text{A22})
$$
\noindent
where $I^{qq}_{\pi,K}(|{\mathbf p}_T|) = I^{q}_{\pi,K}(|{\mathbf p}_T|)$ and $a_{2K}=5$.

\section*{Appendix B} \indent

Here we present some details of our calculation of charged hadron spectra at low $p_T$.
The functions $\Phi_a$ involved into~(\ref{eq:HadronSpectrumStep1a}) and (\ref{eq:HadronSpectrumStep1b}) 
can be presented in the following way:
$$
  \Phi_a(x_\pm, p_T) =  \int\limits_{x_\pm}^{1}dx_1 \int\limits_{0}^{\infty}d{\mathbf k}_{1T}^2 \int\limits_{0}^{2\pi}d{\varphi}_1 \times \nonumber 
$$ 
$$
  \times \int\limits_{x_\pm}^{1-x_1}dx_2 \int\limits_{0}^{\infty}d{\mathbf k}_{2T}^2 \int\limits_{0}^{2\pi}d\varphi_2 {\tilde F}_a(x_1,x_2,{\mathbf k}_{1T}^2,{\mathbf k}_{2T}^2)G_{a\to h}(z,|{\mathbf p}_{T} - z{\mathbf k}_{1T}|),\eqno(\text{B1})
$$
\noindent
where $a = q$, $qq$ or $g$ and $z=x_\pm/x_1$. 
The kernels $\tilde F_a(x_1, x_2, {\mathbf k}_{1T}^2, {\mathbf k}_{2T}^2)$ are
$$
  {\tilde F}_q^g(x_1,x_2,{\mathbf k}_{1T}^2, {\mathbf k}_{2T}^2) = f_q(x_1)g_q({\mathbf k}_{1T}^2)f_{qq}(1-x_1-x_2) g_{qq}(|{\mathbf k}_{1T} + {\mathbf k}_{2T}|^2) f_g(x_2, {\mathbf k}_{2T}^2),\eqno(\text{B2})
$$
$$
  {\tilde F}_{qq}^g(x_1,x_2,{\mathbf k}_{1T}^2, {\mathbf k}_{2T}^2) = f_{qq}(x_1) g_{qq}({\mathbf k}_{1T}^2) f_q(1-x_1-x_2)g_q(|{\mathbf k}_{1T} + {\mathbf k}_{2T}|^2)f_g(x_2, {\mathbf k}_{2T}^2),\eqno(\text{B3})
$$
$$
  {\tilde F}_g(x_1,x_2,{\mathbf k}_{1T}^2, {\mathbf k}_{2T}^2) = f_g(x_1,{\mathbf k}_{1T}^2)f_q(1-x_1-x_2)g_q(|{\mathbf k}_{1T} + {\mathbf k}_{2T}|^2)f_{qq}(x_2) g_{qq}({\mathbf k}_{2T}^2).\eqno(\text{B4})
$$
\noindent
To simplify the integration in~(B1) we perform a change of variables:
$$
  x_1=t_0(1-x_\pm)+x_\pm, \quad {\mathbf k}_{1T}^2=(1-t_1)/t_1, \quad \varphi_1=2\pi t_2, \nonumber 
$$
$$
  x_2=t_3(1-x_1)=t_3(1-x_\pm)(1-t_0)+x_\pm, \quad {\mathbf k}_{2T}^2=(1-t_4)/t_4, \quad \varphi_2=2\pi t_5.\eqno(\text{B5})
$$
\noindent
The integration on all $t$-variables can be now performed in the range $(0, 1)$. The transition Jacobian reads:
$$
J(x_\pm, t_0,...,t_5) = \frac{4\pi^2 (1-x_\pm)(1-x_1(t_0))}{t_1^2 t_4^2}= \frac{4\pi^2 (1-x_\pm)^2(1-t_0)}{t_1^2 t_4^2}.\eqno(\text{B6})
$$
\noindent 

\newpage 

\bibliography{TMD3_GL1}

\end{document}